\documentclass[12pt]{article}
\usepackage{amsmath}
\usepackage{epsfig}
\usepackage{amssymb}
\usepackage{hyperref}
\usepackage{bbold}
\usepackage{braket}
\usepackage{dsfont}

\usepackage{mathtools}
\usepackage{amsfonts}
\usepackage[numbers,sort&compress]{natbib}

\def\be{\begin{equation}}
\def\ee{\end{equation}}
\def\beq{\begin{equation}}
\def\eeq{\end{equation}}
\def\beqa{\begin{eqnarray}}
\def\eeqa{\end{eqnarray}}
\def\ba{\begin{eqnarray}}
\def\ea{\end{eqnarray}}
\def\bea{\begin{eqnarray}}
\def\eea{\end{eqnarray}}

\def\beq{\begin{equation}}
\def\eeq{\end{equation}}
\def\beeq{\begin{eqnarray}}
\def\eeeq{\end{eqnarray}}
\def\to{\rightarrow}
\def\nn{\nonumber}

\def\b0{b_0}

\def\b0{b_0}

\begin{document}

\begin{titlepage}
\renewcommand{\thefootnote}{\fnsymbol{footnote}}
\begin{flushright}
     \end{flushright}
\par \vspace{10mm}
\begin{center}
{\large \bf
T-odd proton-helicity asymmetry in semi-inclusive DIS\\[2mm]
in perturbative QCD}\\

\vspace{8mm}

\today
\end{center}

\par \vspace{2mm}
\begin{center}
{\bf Maurizio Abele${}^{\,a}$,}
\hskip .2cm
{\bf Matthias Aicher${}^{\,b}$,}
\hskip .2cm
{\bf Fulvio Piacenza${}^{\,c}$,}\\[2mm]
\hskip .2cm
{\bf Andreas Sch\"{a}fer${}^{\,b}$,}
\hskip .2cm
{\bf Werner Vogelsang${}^{\,a}$  }\\[5mm]
\vspace{5mm}
${}^{a}\,$ Institute for Theoretical Physics, T\"ubingen University, 
Auf der Morgenstelle 14, \\ 72076 T\"ubingen, Germany\\[2mm]
${}^b$ Institut f\"ur Theoretische Physik, Universit\"at Regensburg,\\
93040 Regensburg, Germany \\[2mm]
${}^c$ Dipartimento di Fisica, Università di Pavia, via Bassi 6,\\ 27100 Pavia, Italy
\end{center}


\vspace{9mm}
\begin{center} {\large \bf Abstract} \end{center}
We compute the single-spin asymmetry $A_{UL}$ in semi-inclusive deep-inelastic scattering of unpolarized leptons and
longitudinally polarized protons at large transverse momentum of the produced hadron. Our calculation is
performed in collinear factorization at the lowest order of QCD perturbation theory. 
For photon exchange the asymmetry is T-odd and receives contributions from 
the interference of the tree level and one-loop absorptive amplitudes. We consider the
behavior of the spin asymmetry at low transverse momentum where contact to the 
formalism based on transverse-momentum dependent distribution functions can be made. 
We also present some phenomenological results relevant for the COMPASS and HERMES experiments
and the future Electron-Ion Collider. 

\end{titlepage}  

\setcounter{footnote}{2}
\renewcommand{\thefootnote}{\fnsymbol{footnote}}

\section{Introduction \label{intro}}

T-odd effects in semi-inclusive deep-inelastic scattering (SIDIS) have been a focus of numerous theoretical and experimental 
studies in recent years~\cite{GrossePerdekamp:2015xdx}. These studies were motivated by the discovery~\cite{Sivers:1989cc,Brodsky:2002cx,Collins:2002kn} 
that a proton can in fact have intrinsic T-odd parton distribution functions, associated with the interplay of transverse polarization of the proton or its partons
with the partonic transverse momenta. Here the term ``T-odd'' refers to a ``naive'' time-reversal operation, which corresponds to ordinary time reversal 
without the interchange of initial and final states of the reaction considered. 

T-odd effects can, however, also be generated in perturbation theory. They are absent at tree level, but 
the seminal papers~\cite{DeRujula:1971nnp,DeRujula:1978bz,Fabricius:1980wg,Korner:1980np,Hagiwara:1981qn,Hagiwara:1982cq} described
how they can arise from absorptive parts of loop amplitudes at ${\cal O}(\alpha_s^2)$ in QCD hard scattering,
where $\alpha_s$ is the strong coupling. Initially proposed as tests of QCD and its gluon 
self-coupling~\cite{DeRujula:1978bz,Fabricius:1980wg,Hagiwara:1981qn}, T-odd effects in perturbative QCD
have remained a subject of interest ever since~\cite{Pire:1983tv,Hagiwara:1984hi,Bilal:1990wi,Carlitz:1992fv,Brandenburg:1995nv,Ahmed:1999ix,Korner:2000zr,Hagiwara:2007sz,Frederix:2014cba,Benic:2019zvg}.
In regards to SIDIS, the early studies~\cite{Hagiwara:1982cq,Ahmed:1999ix,Korner:2000zr} have addressed
neutrino scattering as well as scattering of longitudinally polarized leptons off unpolarized protons~\cite{Ahmed:1999ix,Korner:2000zr}. 

In the present paper, we extend the previous work and compute the leading perturbative T-odd effects for 
SIDIS with unpolarized leptons colliding with longitudinally polarized protons via photon exchange~\cite{aicher,abele,Piacenza:2020sst}, which to our knowledge 
have not been investigated by other authors. Calculating the relevant absorptive parts of one-loop amplitudes,
and using collinear factorization, we derive the corresponding azimuthal terms in the 
spin asymmetry $A_{UL}$ when the proton beam helicity is flipped. 
Our calculation is to be seen in the same spirit as other approaches that aim to obtain the phase
required for (in their case, transverse) single-spin asymmetries through a hard-scattering mechanism~\cite{Qiu:1998ia,Benic:2019zvg}. 

There are several aspects of this observable that motivate us to carry out this study. First, and perhaps foremost,
perturbative T-odd effects in QCD have remained elusive so far, and given their unique property of arising from loop effects in 
QCD, any observable sensitive to them is valuable. In this context it is also worth mentioning that for $A_{UL}$
the effects are sensitive to the proton's helicity parton distributions {\it despite} the fact that an unpolarized
lepton beam is used. This is quite unique as well, since usually conservation of parity in strong
interactions prohibits such single-longitudinal spin asymmetries.

Second, measurements of the relevant azimuthal terms
have been carried out in various fixed-target experiments by the HERMES \cite{HERMES:1999ryv,HERMES:2001hbj,HERMES:2002buj,HERMES:2005mov},
CLAS \cite{CLAS:2010fns} and COMPASS \cite{COMPASS:2010nak,Savin:2016gah,COMPASS:2016klq} collaborations, 
albeit in kinematic regions that are not clearly in the perturbative regime. Nevertheless, it is interesting to see whether
the perturbative calculations give results that are roughly consistent with data at the highest transverse momenta $P_{h\perp}$ of the produced
hadron accessed so far. Much higher $P_{h\perp}$ should become available at the future Electron-Ion Collider (EIC), where
SIDIS studies with exquisite precision will be feasible~\cite{Aschenauer:2019kzf}. It is therefore valuable to extend the
``library'' of observables relevant at the EIC.  

Finally, as mentioned above, most studies of T-odd effects in QCD have addressed the non-perturbative 
regime in terms of parton distributions and fragmentation functions. For SIDIS, this approach becomes particularly
useful when the transverse momentum of the outgoing hadron is relatively low, $P^{2}_{h\perp}\ll Q^2$, with $Q^2$
the virtuality of the exchanged photon. In this case one can describe SIDIS in terms of ``Transverse Momentum Dependent'' (TMD)
parton distributions and fragmentation functions~\cite{Rogers:2015sqa,Angeles-Martinez:2015sea}.  
As has been shown~\cite{Bacchetta:2004zf,Bacchetta:2006tn}, TMDs can indeed generate the SIDIS spin asymmetry $A_{UL}$,
and numerous phenomenological studies have been performed~\cite{Boglione:2000jk,Ma:2000ip,Ma:2001ie,Efremov:2003tf,Boffi:2009sh,Zhu:2011zza,Lu:2011pt,Li:2021mmi}.
Having also a perturbative calculation of $A_{UL}$, for which the observed transverse momentum is acquired by the recoil 
against a hard parton in the scattering process,
one can address the question in how far the TMD formalism is recovered as one
takes the limit $P^{2}_{h\perp}\ll Q^2$. General statements about the high-transverse-momentum tail of TMDs were
developed in~\cite{Bacchetta:2008xw}, which also make predictions for the behavior of $A_{UL}$ that may be directly compared
to our results. In this context also the T-odd beam-spin asymmetry $A_{LU}$ is interesting~\cite{Ahmed:1999ix,Korner:2000zr,Yuan:2003gu,Afanasev:2006gw}, for which 
the initial lepton is polarized, and we will briefly discuss this asymmetry as well.
We note that additional insights into the matching of TMDs to perturbative calculations have become available in recent years~\cite{Ji:2006ub,Ji:2006br,Yuan:2009dw,Kanazawa:2015ajw,Scimemi:2019gge,Moos:2020wvd,Vladimirov:2021hdn,Scimemi:2018mmi,Ebert:2021jhy,Rodini:2022wki}.

Our paper is organized as follows. In Sec.~\ref{pSIDIS} we introduce the kinematic variables and the main ingredients for the 
perturbative description of the spin-dependent SIDIS cross section. In Sec.~\ref{ssa} we briefly review the main 
properties of T-odd asymmetries and describe the strategy for our calculation. Section~\ref{collres} presents 
our perturbative results for the T-odd contributions to the SIDIS spin asymmetry. Next, in Sec.~\ref{sec:LOWQT}, we consider the
limit of small transverse momenta and compare to known results in the TMD regime. 
Phenomenological results are presented in Sec.~\ref{Pheno}. Here we consider the spin asymmetry $A_{UL}$ 
at the EIC and also compare to the COMPASS \cite{COMPASS:2016klq} and HERMES data \cite{HERMES:1999ryv,HERMES:2005mov}. Section~\ref{conc} concludes our paper.

\section{Perturbative SIDIS cross section \label{pSIDIS}}

\begin{figure}[t]
\centering
\includegraphics[scale=0.9]{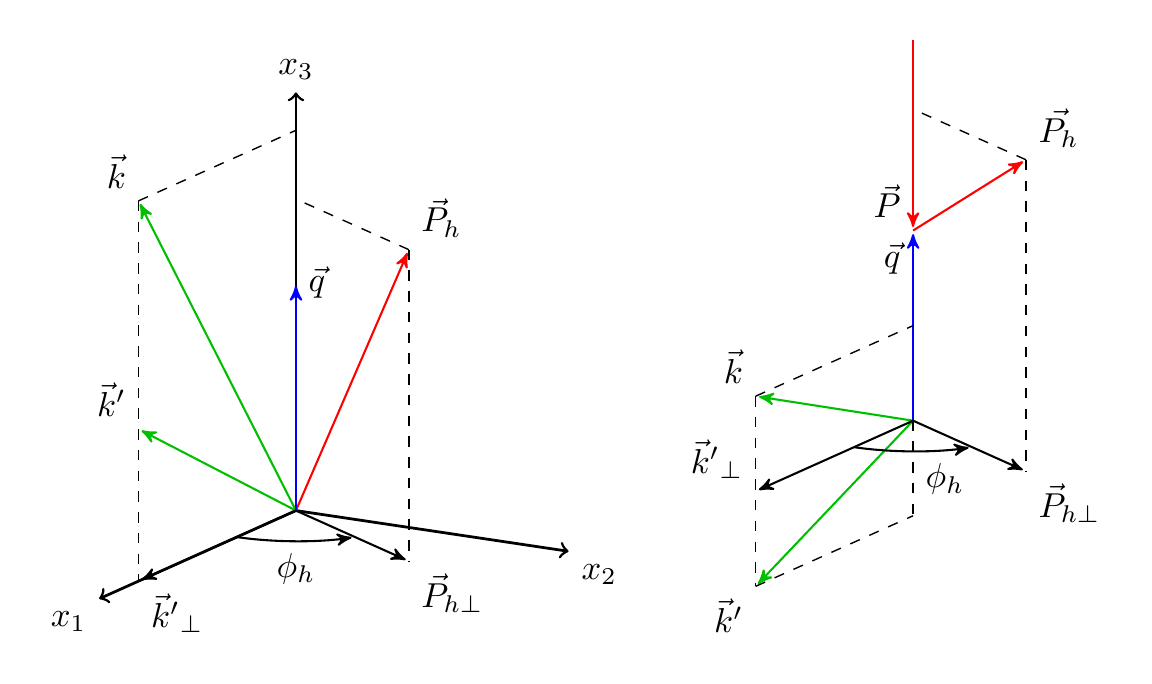}  
\caption{\it{Left: kinematics of the SIDIS process in the rest frame of the proton. Right: same in the Breit frame.}}
\label{fig:kinematics}
\end{figure}

We consider the SIDIS process 
$$
\ell(k) + p(P,S) \to \ell'(k') + h(P_{h}) + X\,,
$$
where we have indicated the four-momenta of the participating particles, and where $S$ is the proton spin vector. 
We set $q \equiv k -k'$ and $Q^2 \equiv -q^2$ for the exchanged virtual gauge boson, for which we will consider only 
 a virtual photon, thus excluding parity-violating effects. The usual kinematical variables relevant for SIDIS are defined as
\begin{equation}
	x = \frac{Q^{2}}{2 P\cdot q}\; , \quad
	y = \frac{P \cdot q}{P\cdot  k}\; , \quad
	z = \frac{P \cdot P_{h}}{P \cdot q}\,.
	\label{eq:DISquant}
\end{equation}
In the following, we will consider the transverse momentum $P_{h\perp}$ and its azimuthal angle $\phi_{h}$ with respect to the lepton plane,
defined in a suitable reference frame. For SIDIS phenomenology one usually adopts the proton rest frame. The kinematics of the process in this frame are 
depicted on the left side of Fig.~\ref{fig:kinematics}. The $x_3$ axis is defined by the direction of the photon three-momentum $\vec{q}$. 
Our actual calculations will be performed in the Breit frame in which the photon four-momentum has a vanishing energy component,
$q= (0,0,0,Q)$, which simplifies the calculations. 
This frame is related to the rest frame by a longitudinal boost along the $x_3$ axis so that all transverse components
remain unchanged. 
The situation in the Breit frame is shown on the right side of Fig.~\ref{fig:kinematics}. 

As discussed in the Introduction, we consider longitudinal polarization for the proton. In the proton rest frame, this
is defined by choosing the proton's spin vector along (or opposite to) the direction of the virtual photon.
Here $\vec{S} \parallel \vec{q}$ will correspond to negative longitudinal polarization of the proton.
We note that in actual experiments one will define longitudinal polarization in the proton rest frame by choosing the spin parallel or
antiparallel to the {\it lepton} beam direction rather than the photon one. 
The two cases are, of course, related; all details may be found in Ref.~\cite{Diehl:2005pc} (see also~\cite{HERMES:2005mov}).
Specifically, they differ by admixtures related to the corresponding transverse single-spin asymmetry $A_{UT}$,
which can be taken into account in the experimental analysis. 
Note that the case with polarization along the lepton beam direction readily extends to the situation at an
$\ell p$ collider, where a longitudinally polarized proton will be in a helicity state. In the following we will therefore
consider protons with either positive or negative helicity. 

For an incoming unpolarized lepton scattering off a longitudinally polarized proton two independent structure functions
contribute to the proton helicity-dependent part of the cross section~\cite{Mendez:1978zx,Bacchetta:2006tn,Bacchetta:2008xw}, 
entering with dependences of the form $\sin(\phi_h)$ or $\sin (2\phi_h)$, respectively. Explicitly, we have
\begin{eqnarray}
\label{spindepcross}
\frac{d^{\,5}\Delta\sigma^h}{dx\; dy\; dz\; dP^{2}_{h\perp}\, d\phi_{h}}&=&
\frac{1}{2}\left(\frac{d^{\,5}\sigma^h_+}{dx\; dy\; dz\; dP^{2}_{h\perp}\,d\phi_{h}} - 
\frac{d^{\,5}\sigma^h_-}{dx\; dy\; dz\; dP^{2}_{h\perp}\,d\phi_{h}} \right)\nn\\[2mm]
	&=&\frac{\pi \alpha^{2}}{x Q^{2}}\frac{y}{1-\varepsilon} 
	\biggl\{
	\sqrt{2\varepsilon (1+ \varepsilon)}\,F^{\,\sin\phi_{h}}_{UL}\sin(\phi_{h})
	+ \varepsilon \,F^{\,\sin2\phi_{h}}_{UL} \sin(2\phi_{h})
	\biggr\}\, ,
\end{eqnarray}
where the subscripts $\pm$ denote proton helicities 
and $\varepsilon$ is defined as the ratio of longitudinal and transverse photon fluxes,
\begin{equation}
	\varepsilon\,
		\equiv\, \frac{1-y}{1-y+y^{2}/2}\, .
\end{equation}
The structure functions $F^{\,\sin\phi_{h}}_{UL},F^{\,\sin2\phi_{h}}_{UL}$ depend on $x,z,Q^2$ and $P^{2}_{h\perp}$,
which we will usually not write out.  In the following, we will compute them
in collinear factorization, where they become double convolutions of helicity parton distribution functions, 
fragmentation functions, and perturbative partonic coefficient functions. We will only
consider the lowest order (LO) in perturbation theory, at which the structure functions may
be cast into the forms
\beeq\label{structfunc}
F^{\,\sin\phi_{h}}_{UL}&=&\left(\frac{ \alpha_{s}(\mu^2)}{2\pi}\right)^2\,\frac{x}{Q^{2}z^2}
\sum_{\begin{subarray}{c}
a,b \\
=\,q,\bar{q},g \end{subarray}} 
\int_{x}^{1} \frac{d\hat{x}}{\hat{x}}
	\int_{z}^{1} \frac{d\hat{z}}{\hat{z}}\,\Delta f_{a}\left(\frac{x}{\hat{x}},\mu^{2}\right)\,
	 		C^{\,\sin\phi_h\,,a\to b}_{UL}(\hat{x}, \hat{z})
		\,D^h_{b}\left(\frac{z}{\hat{z}},\mu^{2}\right)\nn\\[2mm]
		&\times&\;\delta\left( \frac{q_{T}^{2}}{Q^{2}} - \frac{(1 - \hat{x})(1 - \hat{z})}{\hat{x} \hat{z}}\right)\,,
\eeeq
and likewise for $F^{\,\sin2\phi_{h}}_{UL}$. The factor $(\alpha_{s}/(2\pi))^2 \, x/(Q^2z^2)$ has
been introduced for convenience; it explicitly exhibits the leading power of $\alpha_s$ of the structure functions and also
makes the coefficient functions $C^{\,\sin\phi_h,a\to b}_{UL}$, $C^{\,\sin2\phi_h,a\to b}_{UL}$ 
dimensionless functions of only the two partonic variables
\beq
\hat{x} \,\equiv\, \frac{Q^2}{2 p_a\cdot q}\; , \quad
	\hat{z} \,\equiv\, \frac{p_a \cdot p_b}{p_a \cdot q}\,,
\eeq
which are the partonic counterparts of the hadronic variables in Eq.~(\ref{eq:DISquant}). 
The coefficient functions are to be derived for each $2\to 2$ partonic channel $\gamma^*+a\to b+c$, where parton $b$ fragments into
the observed hadron and parton $c$ remains unobserved. These processes are 
$\gamma^* q(\bar{q})\to q(\bar{q})g$, $\gamma^* q(\bar{q})\to gq(\bar{q})$, $\gamma^* g\to q\bar{q}$, and 
$\gamma^* g\to \bar{q}q$. 

In Eq.~\eqref{structfunc} $\Delta f_a(\xi,\mu^2)$ is the helicity distribution of parton $a=q,\bar{q},g$ in the proton at momentum 
fraction $\xi$ and factorization scale $\mu$ (which, for simplicity, we choose equal to the renormalization scale $\mu$ 
appearing in the strong coupling constant $\alpha_s$). 
Furthermore, $D^h_{b} \left(\zeta,\mu^2\right)$ is the corresponding fragmentation function for parton $b$ 
going to the observed hadron $h$, at momentum fraction $\zeta$ and, again, at factorization scale $\mu$. 
All functions in Eq.~\eqref{structfunc} are tied together by the $\delta$ function in the second line which
expresses the fact that at LO the recoiling partonic system consists of a single massless parton $c$. 
For convenience, we have introduced the variable 
\begin{align}
	q^{2}_{T} \equiv \frac{P_{h\perp}^{2}}{z^{2}}\, .
	\label{eq:kqt}
\end{align}

\section{T-odd single-spin asymmetry at lowest order}
\label{ssa}

The terms proportional to $\sin(\phi_h)$ and $\sin(2\phi_h)$ represent correlations of the forms
$\vec{S}\cdot (\vec{k}{'}_\perp\times\vec{P}_{h\perp})$ and $\vec{S}\cdot (\vec{k}{'}_\perp\times\vec{P}_{h\perp})
(\vec{k}{'}_\perp\cdot\vec{P}_{h\perp})$, respectively, which already suggests that they are ``naively'' time-reversal odd. 
This sets a constraint on the partonic scattering processes that may contribute to the corresponding asymmetries
in perturbation theory. To set the stage for our derivations, we briefly review how this constraint can be exploited to simplify the calculations.

\begin{figure}
\centering
\hspace*{-2mm}
\includegraphics[scale=1]{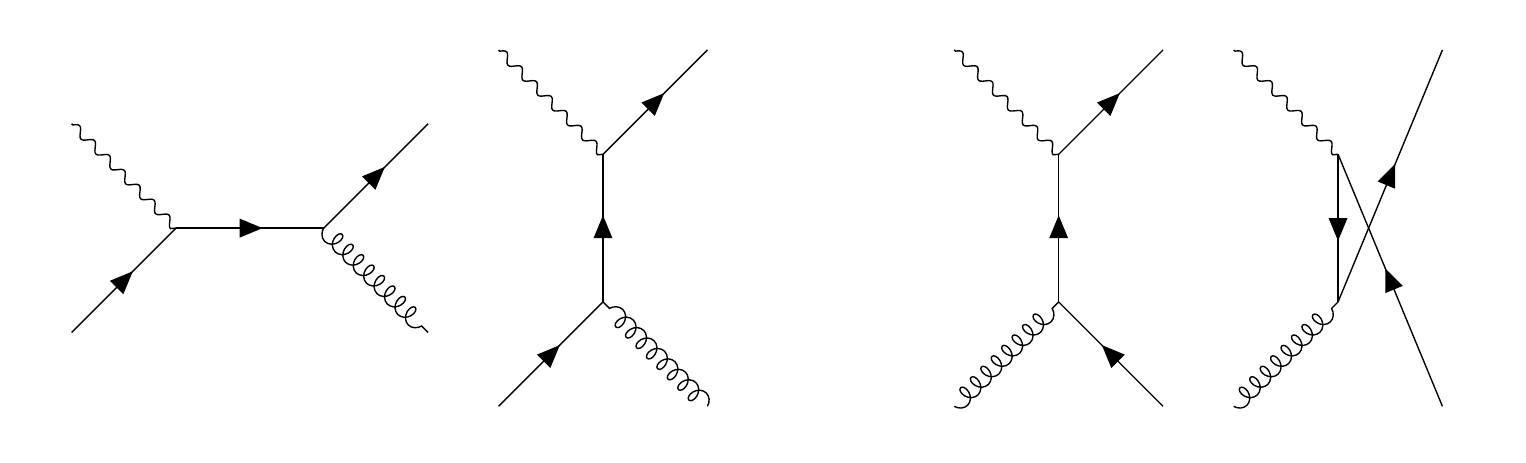}
\vspace*{-2mm}
\caption{{\it Tree-level diagrams for $\gamma^{*} + q \rightarrow q +g$ and $\gamma^{*} + g \rightarrow q+ \bar{q}$.}}
\label{fig:tree-level}
\end{figure}

\begin{figure}
\vspace*{5mm}
\centering
\hspace*{-2mm}
\includegraphics[scale=.9]{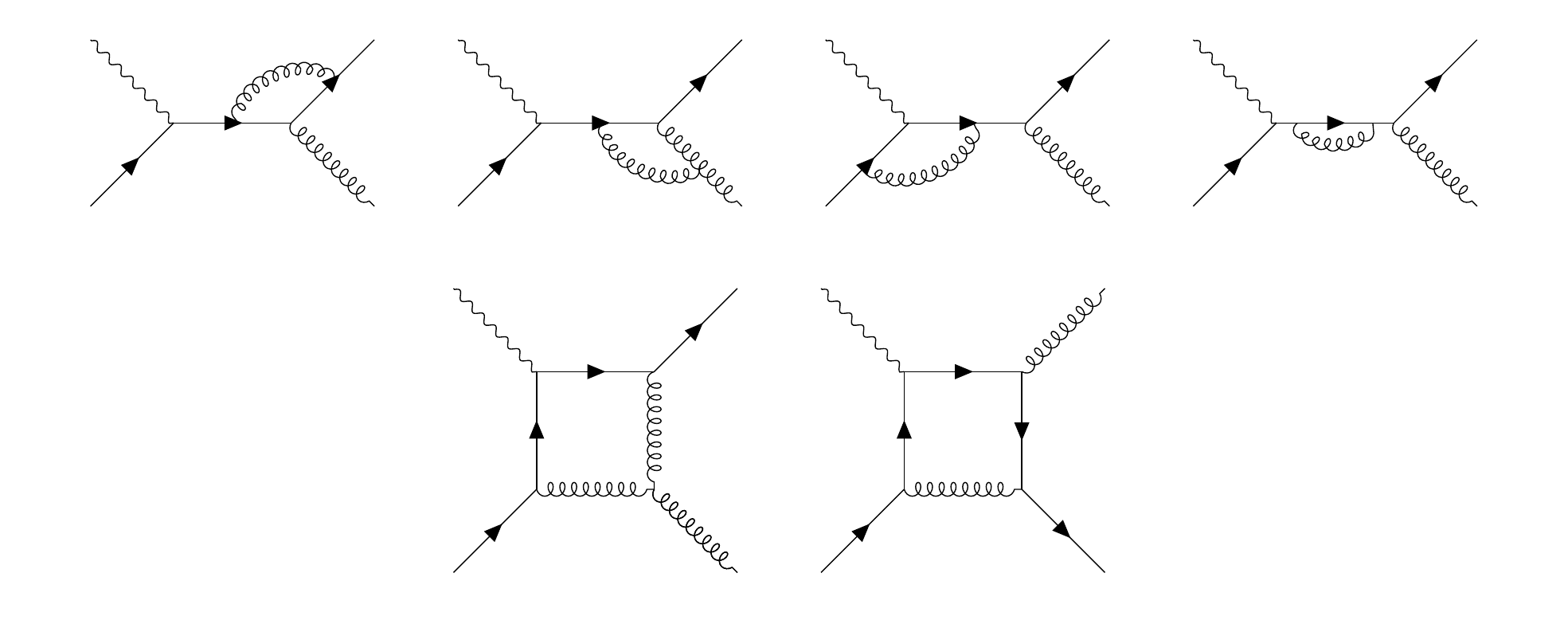}
\caption{{\it One-loop diagrams for $\gamma^{*} + q \rightarrow q +g$ that provide a phase relative to the tree-level
amplitude. We note that there are additional one-loop diagrams that do not produce a phase.}}
\label{fig:quarkloop}
\end{figure}

\begin{figure}
\vspace*{5mm}
\centering
\hspace*{-2mm}
\includegraphics[scale=1]{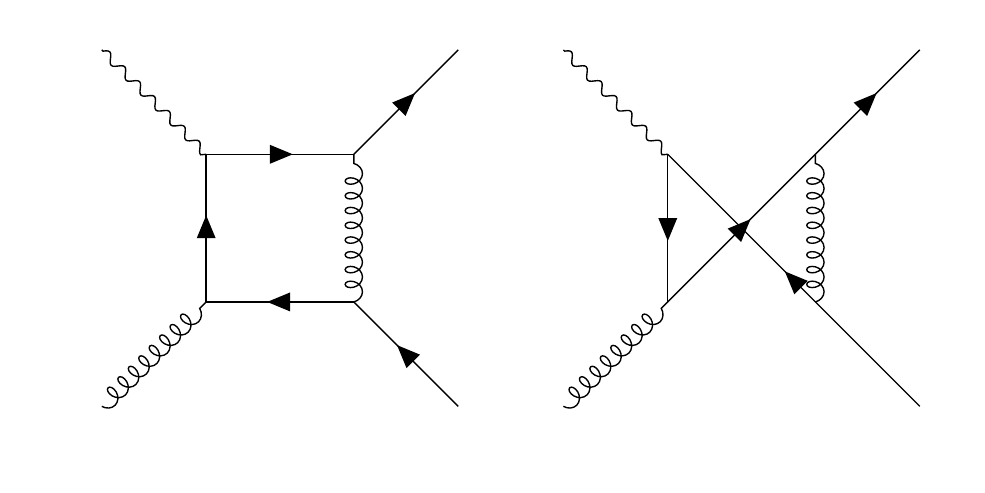}
\caption{{\it One-loop diagrams for $\gamma^{*} + g \rightarrow q +\bar{q}$ that provide a phase relative to the tree-level
amplitude.}}
\label{fig:gluonloop}
\end{figure}

Denoting as $S_{fi}$ the scattering matrix element between an initial state $i$ and a final state $f$, 
a \textit{naive} time-reversal transformation corresponds to a time-reversal without interchange of initial and final states. Hence a T-odd observable is 
characterized by~\cite{Karpman:1968gvz,Cannata:1970br,Hagiwara:1982cq} 
\begin{equation}
\big|S_{fi}\big|^2 \neq \big|S_{\tilde{f}\, \tilde{i}}\big|^2\,,
\end{equation}
where $\tilde{i}$($\tilde{f}$) is obtained from $i$($f$) by reversing momenta and spins. T-odd effects can also be present in theories which are invariant under \textit{true} time-reversal, fulfilling
\begin{equation}
\label{eq:Treversal}
\big|S_{fi}\big|^2 = \big|S_{ \tilde{i}\tilde{f}}\big|^2\,.
\end{equation}
This is easily understood by considering the reaction matrix $T$:
\begin{equation}
S_{fi} \equiv \delta_{fi} + i(2\pi)^4\delta^{(4)}\left(P_f -P_i \right)T_{fi}\,,
\end{equation}
and the unitarity condition for the scattering matrix
\begin{equation}
\label{eq:unitarity}
T_{fi}-T^*_{if} = i \sum_X T^*_{Xf}T_{Xi}\delta^{(4)}\left(P_X - P_i \right) \equiv i \alpha_{fi}\,,
\end{equation}
where in the last equation we introduced the absorptive part $\alpha_{fi}$ of the reaction amplitude. Eq.~\eqref{eq:unitarity} can be rewritten as
$T^*_{if} =T_{fi}-  i \alpha_{fi}$. Taking the square modulus of both sides we find
\begin{equation}
\label{eq:squareUnitarity}
\left|T_{if} \right|^2 = \left|T_{fi} \right|^2+2\, \mathrm{Im} \left(T_{fi}^*\alpha_{fi}  \right)+ \left|\alpha_{fi} \right|^2 \,.
\end{equation}
True time reversal invariance, Eq.~\eqref{eq:Treversal}, implies $\left|T_{if} \right|^2 = \left|T_{\tilde{f}\,\tilde{i}} \right|^2$ (leaving aside the case $i=f$). Thus, if only QED and QCD interactions are present, Eq. \eqref{eq:squareUnitarity}  gives an expression for T-odd terms:
\begin{equation}
\label{eq:squareTodd}
 \big|T_{\tilde{f}\,\tilde{i}} \big|^2 - \big|T_{fi} \big|^2 = 2\, \mathrm{Im} \left(T_{fi}^*\alpha_{fi}  \right)+ \left|\alpha_{fi} \right|^2\,.
\end{equation}
If we consider the partonic processes underlying semi-inclusive DIS, the LO contributions to $T_{fi}$ are the tree-level diagrams for $\gamma^{*} + q \rightarrow q +g$ and $\gamma^{*} + g \rightarrow q +\bar{q}$ shown in Fig.~\ref{fig:tree-level}. The leading terms for the absorptive amplitude $\alpha_{fi} $ arise from 
loop corrections already at one-loop order. The one-loop diagrams shown in Fig.~\ref{fig:quarkloop} for the initial-quark channel and in 
Fig.~\ref{fig:gluonloop} for the initial-gluon channel all have the property that they have an imaginary part and hence produce a
phase relative to the corresponding tree-level amplitudes. As a result, the term $2 \,\mathrm{Im}(\, T^{*}_{fi}\alpha_{fi})$ in Eq.~(\ref{eq:squareTodd})
is non-vanishing already due to the interferences of the one-loop and tree amplitudes. We conclude that LO contributions to T-odd effects in SIDIS come precisely 
from these interferences and are of order $\mathcal{O}(\alpha^{2}_{s})$~\cite{Hagiwara:1982cq}. 

Let us briefly describe the strategy we have adopted in computing the T-odd interference contributions. Introducing
the amplitudes $\mathcal{M}^{\pm}_{ab}$ for positive and negative helicity of parton $a$ in the channel $\gamma^*+a\to b+c$, 
we write the difference of their squares as 
\begin{equation}
	\vert\mathcal{M}^{+}_{ab} \vert^{2} -\vert\mathcal{M}^{-}_{ab} \vert^{2}
		\,=\, L^{\mu\nu} \Big( \hat{W}^{+}_{\mu\nu} -\hat{W}^{-}_{\mu\nu}\Big)\,\equiv\,L^{\mu\nu}\Delta\hat{W}_{\mu\nu} \,,
	\label{eq:scatamp}
\end{equation}
where
\begin{equation}
	L_{\mu\nu}
		= 2 \left(
			 k_{\mu}  k'_{\nu}
			+  k_{\nu}  k'_{\mu}
			- \frac{Q^{2}}{2}g_{\mu\nu}
		\right)
	\label{eq:leptensor}
\end{equation}
is the leptonic tensor, and 
\begin{equation}
	\Delta \hat{W}_{\mu\nu}
		\,\equiv\,{\braket{p_a, +\vert J_{\mu}(0) \vert p_b\, p_c}} \,
		{\braket{p_b\, p_c \vert J_{\nu}(0) \vert p_a, +}}-{\braket{p_a, -\vert J_{\mu}(0) \vert p_b\, p_c}} \,
		{\braket{p_b\, p_c \vert J_{\nu}(0) \vert p_a, -}}
\end{equation}
is the partonic tensor for a polarized parton $a$ in the initial state and a fragmenting parton $b$ in the final state (at LO, the final state
is completely fixed by $a$ and $b$). Since, as discussed above, only interferences between tree-level and one-loop amplitudes
contribute to the order we are considering, we have
\begin{eqnarray}\label{eq:interference} 
&&\hspace*{-1cm}	\Delta \hat{W}_{\mu\nu}\nn\\[2mm]
	&=&\Big[
		{\braket{p_a, +\vert J_{\mu,\text{tree}}(0) \vert p_b\, p_c}} \,
		{\braket{p_b\, p_c \vert J_{\nu,\text{loop}}(0) \vert p_a, +}}+
		{\braket{p_a, +\vert J_{\mu,\text{loop}}(0) \vert p_b\, p_c}} \,
		{\braket{p_b\, p_c \vert J_{\nu,\text{tree}}(0) \vert p_a, +}} 
		\Big]\nn\\[2mm]
		&-&
		\Big[
		{\braket{p_a, -\vert J_{\mu,\text{tree}}(0) \vert p_b\, p_c}} \,
		{\braket{p_b\, p_c \vert J_{\nu,\text{loop}}(0) \vert p_a,-}}+
		{\braket{p_a, -\vert J_{\mu,\text{loop}}(0) \vert p_b\, p_c}} \,
		{\braket{p_b\, p_c \vert J_{\nu,\text{tree}}(0) \vert p_a,-}}
		\Big]\,. \nn\\
\end{eqnarray}
The phase required for a non-vanishing imaginary part is generated by analytic continuation of logarithms in the loop integrals, e.g.,
\begin{equation}
\ln\left( -\frac{\mu^{2}}{\hat{s} +\mathrm{i}\epsilon} \right)
	\quad \longrightarrow \quad
\ln\left( \frac{\mu^{2}}{\hat{s}} \right) + \mathrm{i} \pi\, ,
\end{equation}
where $\hat{s}= (q + p)^{2} = Q^{2}(1-\hat{x})/\hat{x}$. As mentioned, such phases only appear in the $s$-channel diagrams  and the two box-diagrams in Fig.~\ref{fig:quarkloop}, for initial quarks (or antiquarks). For initial gluons, they appear in the two box diagrams in Fig.~\ref{fig:gluonloop}. 

It is quite straightforward to compute the partonic tensor $\Delta \hat{W}_{\mu\nu}$. The only subtlety is related to the use of the 
Dirac matrix $\gamma_5$ and the Levi-Civita tensor $\varepsilon_{\mu\nu\rho\sigma}$ in dimensional regularization in 
$d=4-2\epsilon$ dimensions. Here we have used the scheme of Refs.~\cite{tHooft:1972tcz,Breitenlohner:1977hr} which is
known to be algebraically consistent. We have used the Mathematica package \texttt{Tracer} \cite{Jamin:1991dp} to compute the Dirac traces 
and contractions and \texttt{Package-X} \cite{Patel:2016fam} for the evaluation of the loop integrals and their imaginary parts. 
We note that poles in $1/\epsilon$ arise at various intermediate stages of the calculation; these all have to cancel in the end
since, at the lowest order, the T-odd part of the hadronic tensor must not have any ultraviolet or infrared/collinear singularities.
This provides a useful check on our calculation. The partonic coefficient functions are found from the final result for 
$L^{\mu\nu}\Delta\hat{W}_{\mu\nu}$ as the coefficients of the terms $\sim\sin(\phi_h)$ and $\sim\sin(2\phi_h)$. No
other angular dependences appear.

\section{Analytical results}
\label{collres}

For the partonic coefficient functions for $F^{\,\sin\phi_{h}}_{UL}$ we find~\cite{abele,Piacenza:2020sst}:
\begin{align}
	C^{\,\sin\phi_h,q \to q}_{UL}(\hat{x}, \hat{z})
		&=\, e_q^2 \,C_{F}
			\biggl( C_{A}(1-\hat{x}) + C_{F}(\hat{x}-1-\hat{z} +3\hat{x}\hat{z})\nn\\[2mm]
		&\left.+\, (C_{A}-2C_{F})(1-2\hat{x})\frac{\hat{z}\ln\hat{z}}{1-\hat{z}} 
			\right) \frac{Q}{q_T}\, ,\nn\\[2mm]
C^{\,\sin\phi_h,q\to  g}_{UL}(\hat{x}, \hat{z})
		&=\, -e_q^2 \,C_F \frac{(1-\hat{z})}{\hat{z}} \biggl( C_A (1-\hat{x})+C_F (-3 \hat{x} \hat{z}+4 \hat{x}+\hat{z}-2)  \nn\\[2mm]
		&\left.	+ \,(C_A - 2 C_F)(1-2 \hat{x})\frac{ (1-\hat{z})  \ln (1-\hat{z})}{\hat{z}}\right)\frac{Q}{q_T} \, ,\nn\\[2mm]
C^{\,\sin\phi_h, g\rightarrow q }_{UL}(\hat{x}, \hat{z})
		&=\, e_q^2\,(C_A-2 C_F)(1-\hat{x})\frac{1}{2\hat{z}^2}\biggl(\hat{x} \hat{z} (1-2 \hat{z})-(1-\hat{x}) \ln (1-\hat{z})\nn\\[2mm]
		&\left.	+\,(1-\hat{x})\frac{ \hat{z} \ln (\hat{z})}{1-\hat{z}} \right)\frac{Q}{q_T}\, ,
		\label{CULphi}
\end{align}
while for the coefficients for $F^{\,\sin2\phi_{h}}_{UL}$ we have~\cite{abele,Piacenza:2020sst}
\begin{align}
	C^{\,\sin2\phi_h, q\rightarrow q}_{UL}(\hat{x}, \hat{z})
		&=\, e_q^2\, C_{F} (1-\hat{x})\left(
			(C_A-2 C_F)\frac{(1-2 \hat{z})  \ln\hat{z}}{1-\hat{z}} 
			-(C_A+(1-3 \hat{z})C_F)\right)\frac{Q^{2}}{q^{2}_T}\, ,\nn\\[2mm]
C^{\,\sin2\phi_h,q\rightarrow g}_{UL}(\hat{x}, \hat{z})
		&=\, -e_q^2\, C_{F} (1-\hat{x})\frac{(1-\hat{z})^2}{\hat{z}^2}\left( (C_A-2 C_F) \frac{(1-2 \hat{z})\ln (1-\hat{z})}{\hat{z}}\right.\nn\\[2mm]
		&+\,(C_A - (2-3 \hat{z})C_F)\biggr)\frac{Q^{2}}{q^{2}_T}\, ,\nn\\[2mm]
C^{\,\sin2\phi_h, g\rightarrow q }_{UL}(\hat{x}, \hat{z})
		&=\, e_q^2\, (C_A-2 C_F)(1-\hat{x})^2\frac{1}{2 \hat{z}^3}\biggl( \hat{z}(2 (1-\hat{z}) \hat{z}-1)-(1-\hat{z}) \ln (1-\hat{z})\nn\\[2mm]
		&\left.	-\,\frac{\hat{z}^2\ln \hat{z} }{1-\hat{z}}\right)\frac{Q^{2}}{q^{2}_T}\,.
	\label{CUL2phi}
\end{align}
Note that the ratio $q_T/Q$ in the above expressions is fixed by $\hat{x}$ and $\hat{z}$ through the $\delta$-function in~(\ref{structfunc}).
The $q\to q$ and $q\to g$ coefficients are related by crossing symmetry in the following way:
\begin{align}
C^{\,\sin\phi_h, q\to g}_{UL}(\hat{x}, \hat{z})
&= -C^{\,\sin\phi_h, q\rightarrow q}_{UL}(\hat{x}, 1-\hat{z})\,, \nn \\[2mm]
C^{\,\sin2\phi_h, q\to g}_{UL}(\hat{x}, \hat{z})
&= C^{\,\sin2\phi_h, q\rightarrow q}_{UL}(\hat{x}, 1-\hat{z})\, .
\label{cross}
\end{align}
Furthermore, because of charge conjugation and parity invariance the results for the antiquark channels $\bar{q}\to\bar{q}$ and
$\bar{q}\to g$ are identical to those for $q\to q$ and $q\to g$, respectively. In addition, we have
\begin{align}
C^{\,\sin\phi_h,  g\rightarrow \bar{q}}_{UL}(\hat{x}, \hat{z})
&= -C^{\,\sin\phi_h, g\rightarrow q}_{UL}(\hat{x}, 1-\hat{z})\,, \nn \\[2mm]
C^{\,\sin2\phi_h, g\rightarrow \bar{q}}_{UL}(\hat{x}, \hat{z})
&= C^{\,\sin2\phi_h, g\rightarrow q}_{UL}(\hat{x}, 1-\hat{z})\, .
\label{qbarg}
\end{align}
In Ref.~\cite{aicher} we have also derived these results via crossing of the corresponding T-odd asymmetries in
$e^{+}e^{-}$~annihilation in~\cite{Korner:2000zr}. Our results given above correct the sign of the results in~\cite{aicher}.
For the case of quark-initiated processes, an independent cross check is provided by the SIDIS beam asymmetries $A_{LU}$ calculated 
in Ref.~\cite{Hagiwara:1982cq}. Indeed, for the charged-current case considered in these papers, the interaction 
mediated by W-bosons selects left-handed quarks, so that even if the target is unpolarized, the partonic matrix elements are the same as 
in our Eq.~\eqref{eq:interference}, albeit with reversed helicity. For instance, by looking at the functions $F_{8}$ and $F_{9}$ in 
Eqs.~(3.14) of \cite{Korner:2000zr}, in the case of quark-initiated diagrams, one can verify that they correspond to 
our $C^{\,\sin\phi_h}_{UL}$ and $C^{\,\sin2\phi_h}_{UL}$ functions with a reversed sign. 
Clearly, this reasoning does not allow comparisons in the case of incoming gluons.

\section{Low-$q_T$ limit}\label{sec:LOWQT}

As discussed in the Introduction, it is interesting to expand the above results for the structure functions 
for low values of $q_T/Q$, in order to make contact with results predicted by the TMD formalism.  
When $q^{2}_{T}/Q^{2}\rightarrow 0$ we can expand the delta condition in~(\ref{structfunc}) as described, for example, 
in \cite{Meng:1991da,Meng:1995yn,Boer:2006eq}:
\begin{align}
	 \frac{1}{\hat{x} \hat{z}}\,\delta &\left(\frac{q_{T}^{2}}{Q^{2}} - \frac{(1-\hat{x})(1-\hat{z})}{\hat{x} \hat{z} } \right)
	 	= \delta(1-\hat{z})\delta(1-\hat{x}) \ln\left(\frac{Q^{2}}{q_{T}^{2}}\right)
	 	+ \frac{\delta(1-\hat{z})}{(1-\hat{x})_{+}}
		+ \frac{\delta(1-\hat{x})}{(1-\hat{z})_{+}}\,+\,{\cal O}\left(\frac{q_T^2}{Q^2}\right)\,.
		\label{eq:deltaexpansion}
\end{align}
To simplify notation, we write the double convolution integral as (we
omit the scale dependence of the parton distributions and fragmentation functions here):
\beq\label{dconv}
\big(\Delta f\otimes {\cal C}\otimes D \big)(x,z)\,\equiv\, 
\int_{x}^{1} \frac{d\hat{x}}{\hat{x}}
	\int_{z}^{1} \frac{d\hat{z}}{\hat{z}}
	\;\delta\left( \frac{q_{T}^{2}}{Q^{2}} - \frac{(1 - \hat{x})(1 - \hat{z})}{\hat{x} \hat{z}}\right)
	\,\Delta f\left(\frac{x}{\hat{x}}\right)\,{\cal C}(\hat{x}, \hat{z})
\, D\left(\frac{z}{\hat{z}}\right)\,.
\eeq  
From the $\sin(\phi_h)$ terms in~(\ref{CULphi}) we then find the following contribution to the $q\to q$ coefficient at low $q_T/Q$:
\begin{align}\label{smallqt1a}
&\hspace*{-10mm}\big(\Delta f_q\otimes C^{\,\sin\phi_h, q\rightarrow q}_{UL} \otimes D_q^h \big)(x,z)\nn \\[2mm]
	&=\,e_q^2\,\frac{Q}{q_{T}}\frac{C_A}{2}\left\{
	-C_F\left( 2 \,\ln\left( \frac{q_T^{2}}{Q^{2}}\right)+3\right) 
	\Delta f_{q}\left(x \right)	D^h_{q}\left(z \right) \right.\nn\\[2mm]
	&+ \,D^h_{q}\left(z\right)\int_{x}^{1}d\hat{x}\;
			\delta P_{qq}(\hat{x})\,
			\Delta f_{q}\left(\frac{x}{\hat{x}} \right)+\left. \Delta f_{q}\left(x\right) 
			\int_{z}^{1} d\hat{z}\;\delta P_{qq}(\hat{z})\,D^h_{q}\left(\frac{z}{\hat{z}} \right)
			\right\}\nn\\[2mm]
			&-\,e_q^2\,\frac{Q}{q_{T}}\frac{C_F}{C_A}\,
			\Delta f_{q}\left(x\right) 
			\int_{z}^{1} d\hat{z}\,
			\frac{\hat{z}}{1-\hat{z}}\,
			\left( 1 + \frac{\ln \hat{z}}{1-\hat{z}}\right)
			D^h_{q}\left(\frac{z}{\hat{z}} \right)\,,
\end{align}
where
\beq\label{transv}
\delta P_{qq}(x)\,\equiv\,C_F \left[\frac{2x}{(1-x)_+} + \frac{3}{2} \,\delta(1-x)\right]
\eeq
is the LO splitting function for the scale evolution of the transversity distributions. In the 
$q\to g$ channel we obtain
\begin{align}\label{smallqt1}
&\hspace*{-10mm}\big(\Delta f_q\otimes C^{\,\sin\phi_h, q\rightarrow g}_{UL} \otimes D_g^h \big)(x,z)\nn \\[2mm]
	&=e_q^2\,\frac{Q}{q_{T}}
C_F\,\Delta f_{q}\left(x\right) 
			\int_{z}^{1} d\hat{z}\,
			\frac{1-\hat{z}}{\hat{z}^2}\,
			\big((C_A-2 C_F) \ln(1-\hat{z})-2 C_F \hat{z}\big)
			D_g^h\left(\frac{z}{\hat{z}} \right)\,.
\end{align}
For the process $\gamma^{*} g \rightarrow q \bar{q}$ the coefficient function diverges logarithmically for $\hat{z}\rightarrow 1$. In addition to the 
expansion~(\ref{eq:deltaexpansion}) we therefore also need 
\begin{eqnarray}\label{exp2}
\frac{\ln(1-\hat{z})}{\hat{x}\hat{z}}\,\delta \left(\frac{q_{T}^{2}}{Q^{2}} - \frac{(1-\hat{x})(1-\hat{z})}{\hat{x} \hat{z} } \right)
	\,=\,- \frac{1}{2} \ln^{2}\left(\frac{Q^2}{q_T^{2}}\right)\delta(1-\hat{x})\delta(1-\hat{z})-\frac{\delta(1- \hat{z})}{(1-\hat{x})_{+}}\ln\left(\frac{Q^2}{q_T^{2}}\right)  \nn\\[2mm]
	&&\hspace*{-11cm}
	+\,\delta(1-\hat{x})\left(\frac{\ln(1-\hat{z})}{1-\hat{z}}\right)_{+}-\delta(1- \hat{z})\left(\left(\frac{\ln\left(1-\hat{x}\right)}{1-\hat{x}}\right)_{+} - \frac{\ln\hat{x}}{1-\hat{x}}
	\right)\,.\nn\\
\end{eqnarray}
We then find
\begin{align}\label{smallqt2}
&\hspace*{-2mm}\big(\Delta f_g\otimes C^{\,\sin\phi_h, g\rightarrow q}_{UL} \otimes D_q^h \big)(x,z)]\nn \\[2mm]
	&=\,\frac{e_q^2}{2}\,\frac{Q}{q_{T}}
	(C_A-2 C_{F})D^h_{q}\left(z\right)
	\int_{x}^{1}d\hat{x}\;
\Delta f_{g}\left(\frac{x}{\hat{x}} \right)\left[ (1-\hat{x})\ln\left(\frac{Q^2}{q_T^{2}}\right)+ (1-\hat{x}) \ln\left(\frac{1-\hat{x}}{\hat{x}}\right) -1  \right]\,.
\end{align}
The results in Eqs.~(\ref{smallqt1a}),(\ref{smallqt1}),(\ref{smallqt2}) are valid up to terms of order $q_T/Q$. Keeping in mind the overall
factor $1/Q^2$ in Eq.~(\ref{structfunc}), we see that the structure function $F^{\,\sin\phi_h}_{UL}$
is predicted to have the leading power
\beq\label{Fp1}
F^{\,\sin\phi_h}_{UL}\,\propto\,\frac{1}{Qq_T}\,+\,{\cal O}\left(\frac{1}{Q^2}\right)
\eeq
at low $q_T$, modulo logarithms. The behavior found for  $\gamma^{*} q \rightarrow qg$ in Eq.~(\ref{smallqt1a}) is 
particularly interesting. The term $-C_F(2 \ln(q_T^2/Q^2) +3)$ is the well-known  first-order contribution
to the Sudakov form factor. The next two terms both contain the LO transversity splitting function $\delta P_{qq}$,
convoluted with either the helicity parton distribution or the fragmentation function. A generic low-$q_T$ structure with 
the Sudakov form factor and splitting functions is familiar from the spin-averaged case (see Ref.~\cite{Boer:2006eq}).
However, the appearance of the transversity splitting function in combination with $\Delta f_q$ or $D_q^h$, and along
with an overall factor $C_A$, is quite remarkable. This feature must be 
related to the fact that in the TMD framework the leading part of $A^{\sin\phi_{h}}_{UL}$ receives contributions
from the T-even function $h_L$, which is twist-three and describes the distribution of transversely polarized quarks
in a longitudinally polarized hadron, convoluted with the Collins function probing the fragmentation of transversely polarized
quarks~\cite{Bacchetta:2006tn}. The last term in~(\ref{smallqt1a}) and the results in Eqs.~(\ref{smallqt1}),(\ref{smallqt2}) 
do not appear to have a straightforward structure. Another striking feature is the appearance of a logarithm of $q_T/Q$
in the result for the $g\to q$ channel in Eq.~(\ref{smallqt2}): such logarithms do not usually appear in off-diagonal
contributions at lowest order.

Similarly, we can consider the low-$q_{T}/Q$ limit for the $\sin(2\phi_{h})$ terms. Here we find for the $q\to q$ channel:
\begin{align}\label{smallqt3a}
&\big(\Delta f_q\otimes C^{\,\sin2\phi_h, q\rightarrow q}_{UL} \otimes D_q^h \big)(x,z)\nn \\[2mm]
	&=\,-e_q^2\,\frac{3}{4}\,C_{A}\left\{
	-C_F\left( 2 \,\ln\left( \frac{q_T^{2}}{Q^{2}}\right)+3\right)  \Delta f_{q}\left(x \right) D^h_{q}\left(z \right)\ln\left( \frac{q_{T}^{2}}{Q^{2}}\right) \right.\nn\\[2mm]
&\left.+\,D^h_{q}\left(z\right)\int_{x}^{1}d\hat{x}\;
			\delta P_{qq}(\hat{x})\,
			\Delta f_{q}\left(\frac{x}{\hat{x}} \right)+\Delta f_{q}\left(x\right)\int_{z}^{1}d\hat{z}\;
			\delta P_{qq}(\hat{z})
			D^h_{q}\left(\frac{z}{\hat{z}} \right)	\right\}\nn\\[2mm]
	&+\,\frac{C_F}{2C_A}\,\Delta f_{q}\left(x\right) \,
			\int_{z}^{1} d\hat{z}\,
			\frac{\hat{z}}{(1-\hat{z})^2}\left( 1-3\hat{z}-2 (2\hat{z}-1) 
			\frac{\ln\hat{z}}{1-\hat{z}} \right)
			D^h_{q}\left(\frac{z}{\hat{z}} \right)\,.
\end{align}
Apart from normalization, the first three terms are identical to the corresponding ones in Eq.~(\ref{smallqt1a}). 
We note that, despite first appearances, the integrand of the last term is regular as $\hat{z}\to 1$. 

In the $q\to g$ channel we have
\begin{align}\label{smallqt3}
&\big(\Delta f_q\otimes C^{\,\sin2\phi_h,q\rightarrow g}_{UL} \otimes D_g^h \big)(x,z)\nn \\[2mm]
	&=-e_q^2\,C_{F} \Delta f_{q}\left(x\right) 
			\int_{z}^{1} \frac{d\hat{z}}{\hat z}\,
			\left( (C_A-2 C_F) \frac{(1-2 \hat{z})\ln (1-\hat{z})}{\hat{z}}+(C_A - (2-3 \hat{z})C_F)\right)
			D^h_g\left(\frac{z}{\hat{z}} \right)\,.
\end{align}
For the channel $\gamma^{*} g \rightarrow q \bar{q}$ we again need the expansion~(\ref{exp2}) and obtain
\begin{align}\label{smallqt4}
&\hspace*{-10mm}\big(\Delta f_g\otimes C^{\,\sin2\phi_h, g\rightarrow q}_{UL} \otimes D_q^h \big)(x,z)\nn \\[2mm]
	&=\,\frac{e_q^2}{2}\,(C_A-2 C_{F})D^h_{q}\left(z\right)
	\int_{x}^{1}d\hat{x}\;
			\Delta f_{g}\left(\frac{x}{\hat{x}} \right)\,\hat{x}\,\left[ \ln\left(\frac{Q^2}{q_T^{2}}\right) + \ln\left(\frac{1-\hat{x}}{\hat x}\right) +  \frac{3}{2} \right]\,.
\end{align}
The results in Eqs.~(\ref{smallqt3}),(\ref{smallqt4}) receive corrections of order $q_T^2/Q^2$, so that
\beq\label{Fp2}
F^{\,\sin2\phi_h}_{UL}\,\propto\,\frac{1}{Q^2}\,+\,{\cal O}\left(\frac{q_T^2}{Q^4}\right)\,,
\eeq
again up to logarithms.

Our low-$q_T$ expansions fill two of the gaps reported in Table~2 of Ref.~\cite{Bacchetta:2008xw} providing
the missing perturbative expressions for the $\phi_h$-dependent T-odd cross sections. From the point
of view of TMD factorization, they correspond to the leading part of the ``high-$q_T$ calculation''. As
discussed in~\cite{Bacchetta:2008xw}, the TMD framework predicts the same behavior $\propto 1/(Qq_T)$ 
of the $\sin(\phi_h)$ terms as we find in Eq.~(\ref{Fp1}). In this sense the TMD
calculation matches the collinear one. At this point, however, one cannot decide whether this matching is 
really quantitative in the sense that not just the overall power counting matches, but also
the full combination of hard-scattering coefficients, parton distributions and fragmentation functions.
Currently, despite enormous progress in recent 
years~\cite{Ji:2006ub,Ji:2006br,Yuan:2009dw,Kanazawa:2015ajw,Scimemi:2019gge,Moos:2020wvd,Vladimirov:2021hdn,Scimemi:2018mmi,Ebert:2021jhy,Rodini:2022wki},
the high-transverse-momentum tails of TMDs are not understood at a sufficient level to obtain definitive
results for $A_{UL}$, especially in the case of the fragmentation correlators. Clearly, it will
be very interesting to explore this issue more deeply in the future, also in the light of TMD
factorization theorems extending beyond leading twist proposed recently~\cite{Vladimirov:2021hdn,Ebert:2021jhy}.

For the $\sin(2\phi_h)$ terms, we find in Eq.~(\ref{Fp2}) that the 
perturbative structure function becomes constant at low $q_T$. This is not in accordance
with the TMD prediction~\cite{Bacchetta:2008xw} that the high-$q_T$ tail of the structure function 
behaves as $1/q_T^4$. In the TMD framework, $F^{\,\sin2\phi_h}_{UL}$ is leading twist,
being a convolution of the longitudinal worm-gear functions with Collins functions~\cite{Bacchetta:2006tn}.  

In the context of this discussion, it is also interesting to recall the corresponding results for the T-odd
beam spin asymmetry $A_{LU}$~\cite{Ahmed:1999ix,Korner:2000zr}, for which 
the initial lepton is polarized. The relevant results are given in the Appendix. Interestingly,
at low $q_T$, the same features as described above for $A_{UL}$ are encountered.

\section{Phenomenological results \label{Pheno}}

We now present some simple phenomenology of the T-odd effects in SIDIS with longitudinally
polarized protons. We will not carry out any full-fledged study; rather, we wish to explore the overall size of the 
$\sin\phi_{h}$ and $\sin(2\phi_{h})$ modulations.

The quantities of interest in polarization experiments are typically spin asymmetries. In the present case, 
the longitudinal proton helicity single spin asymmetry in SIDIS is defined as
\begin{equation}
	A_{UL}(\phi_{h})
		\equiv \frac{d\sigma^{h}_+(\phi_{h}) - d\sigma^{h}_{-}(\phi_{h})}{d\sigma^{h}_+(\phi_{h}) + d\sigma^{h}_-(\phi_{h})}\,,
	\label{eq:AUL}
\end{equation}
where as in Sec.~\ref{pSIDIS} $d\sigma^{h}_{\pm}$ represents the (differential) cross section for positive (negative) proton helicity.
The denominator of the asymmetry is just twice the spin-averaged cross section as a function of the azimuthal angle
$\phi_h$. As is well known~\cite{Meng:1991da,Bacchetta:2006tn}, this cross section has a $\phi_h$ independent piece as well
as terms proportional to $\cos(\phi_h)$ and $\cos(2\phi_h)$. Dividing numerator and denominator by the $\phi_h$ independent term, we may write
\begin{equation}
	A_{UL}(\phi_{h})
		=\frac{A^{\sin\phi_{h}}_{UL} \sin\phi_{h} + A^{\sin2\phi_{h}}_{UL} \sin2\phi_{h}}{1 + A^{\cos\phi_{h}}_{UU} \cos\phi_{h} + 
		A^{\cos2\phi_{h}}_{UU} \cos2\phi_{h}}\,.
		\label{eq:AULbis}
\end{equation}
The various angular modulations $A^{\sin\phi_{h}}_{UL}$ etc. are also known as analyzing powers. 
The ones of interest to us here, $A^{\sin\phi_{h}}_{UL}$ and $A^{\sin2\phi_{h}}_{UL}$, may be extracted
from the full cross section as follows:
\begin{equation}
	A^{\sin n\phi_{h}}_{UL}
		= \frac{\int_0^{2\pi} d\phi_{h} \sin(n\phi_{h})\big[d\sigma^{h}_+(\phi_{h}) - d\sigma^{h}_-(\phi_{h})\big] }
		{\frac{1}{2}\int_0^{2\pi}  d\phi_{h}\,\big[d\sigma^{h}_+(\phi_{h}) + d\sigma^{h}_-(\phi_{h})\big] }
		\quad\quad (n=1,2)\,.
		\label{eq:analyzingpow}
\end{equation}
In this way, the terms with $\cos(\phi_h)$ and $\cos(2\phi_h)$ in the spin-averaged cross section do not contribute. 
Experimental data are commonly reported in terms of the $A^{\sin n\phi_{h}}_{UL}$, 
and accordingly these are the quantities that we will consider for our numerical predictions. 

As stated earlier, in the present paper we restrict ourselves to LO predictions for the T-odd terms, keeping
the leading contribution $\propto \alpha_s^2$ in the numerator. For consistency, we therefore also 
need to use only the LO term in the denominator, which is only of order $\alpha_s$ and is
easily computed~\cite{Mendez:1978zx}.  (We note that the NLO
corrections for the spin-averaged cross section in the denominator are available~\cite{Daleo:2004pn,Kniehl:2004hf,Wang:2019bvb}.) 
Because of this mismatch of perturbative orders in the numerator and denominator, 
the analyzing powers $A^{\sin\phi_{h}}_{UL},A^{\sin2\phi_{h}}_{UL}$ are themselves of order $\alpha_s$,
which is in contrast to most other spin asymmetries for which the leading power of $\alpha_s$ cancels. We
therefore expect $A^{\sin\phi_{h}}_{UL},A^{\sin2\phi_{h}}_{UL}$ to be quite sensitive to the choice
of scale and to higher-order corrections. 

For our numerical studies we use the DSSV~\cite{deFlorian:2008mr,deFlorian:2009vb} set for the helicity 
parton distributions and the DSS14 \cite{deFlorian:2014xna} set of fragmentation functions. 
We note that only pion fragmentation is considered in this set. We set the renormalization and factorization scales 
equal to $Q$. For the denominator of the asymmetries we use the NNPDF31 \cite{NNPDF:2017mvq} set of unpolarized parton
distributions. We call this set from the LHAPDF library~\cite{Buckley:2014ana}.

\begin{figure}
\includegraphics[scale=0.75]{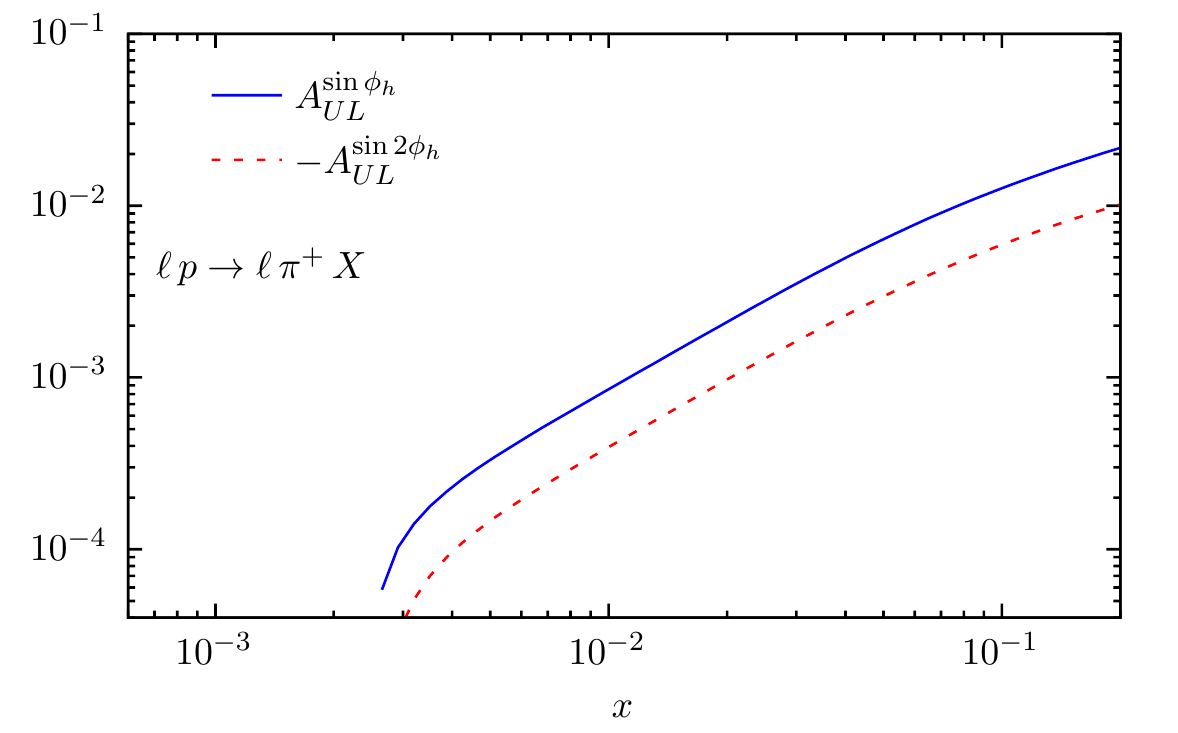}

\vspace*{-5.68cm}
\hspace*{8.6cm}
\includegraphics[scale=0.75]{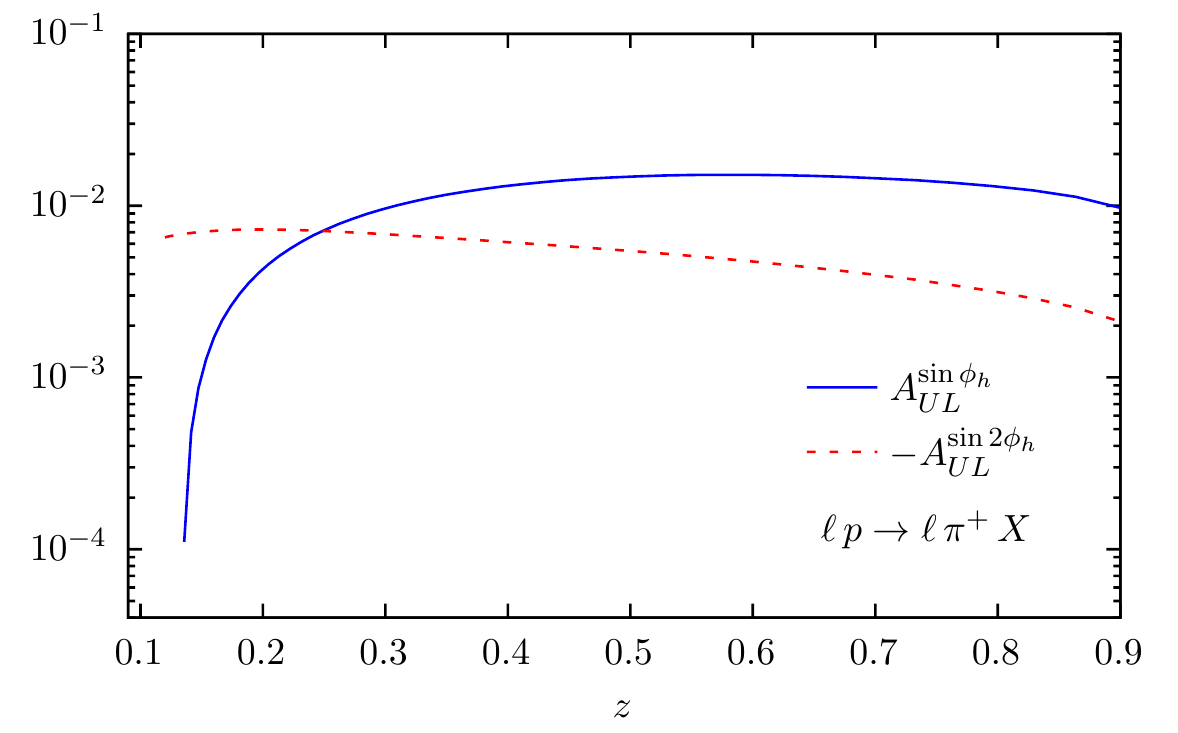}

\caption{\it{Left: $x$-dependence and  Right: $z$-dependence of the analyzing powers for $\ell p\,\rightarrow \ell\, \pi^{+}\, X$ at the future EIC. The $\sin(2\phi_{h})$ analyzing power has been multiplied by $(-1)$. For the left graph, $z=0.4$, while for the right graph $x= 0.1$. In both cases we use $\sqrt{s} = 140\; \mathrm{GeV} $, $Q^{2}=50\;\mathrm{GeV}^{2}$ and $P_{h \perp}=2\;\mathrm{GeV}$.}}
\label{fig:x-and-z-dep}
\end{figure}

We start by presenting estimates for the future Electron-Ion Collider (EIC) with a center-of-mass energy of 140 GeV. 
At this fairly high energy, the $\sin\phi_{h}$ and $\sin(2\phi_{h})$ modulations are overall quite strongly suppressed. The reason is
that at high energies rather low momentum fractions in the parton distribution functions are probed, where
the polarized distributions are much smaller than the unpolarized ones. The left part of Fig.~\ref{fig:x-and-z-dep} shows 
$x$-dependence of the analyzing powers $A^{\sin\phi_{h}}_{UL}$ (blue solid line) and $-A^{\sin2\phi_{h}}_{UL}$ (red dashed line)
for $\pi^+$ production, at 
a set of fixed values of $z$, $Q^{2}$ and $P_{h \perp}$. These values have been chosen by considering the ``projected EIC data'' shown 
in Ref.~\cite{Aschenauer:2019kzf}. We observe that the asymmetries indeed rapidly decrease toward low values of $x$. 
The right part of the figure shows the $z$-dependence of the asymmetries, which is much more moderate through most of the range
considered.

\begin{figure}
\includegraphics[scale=0.75]{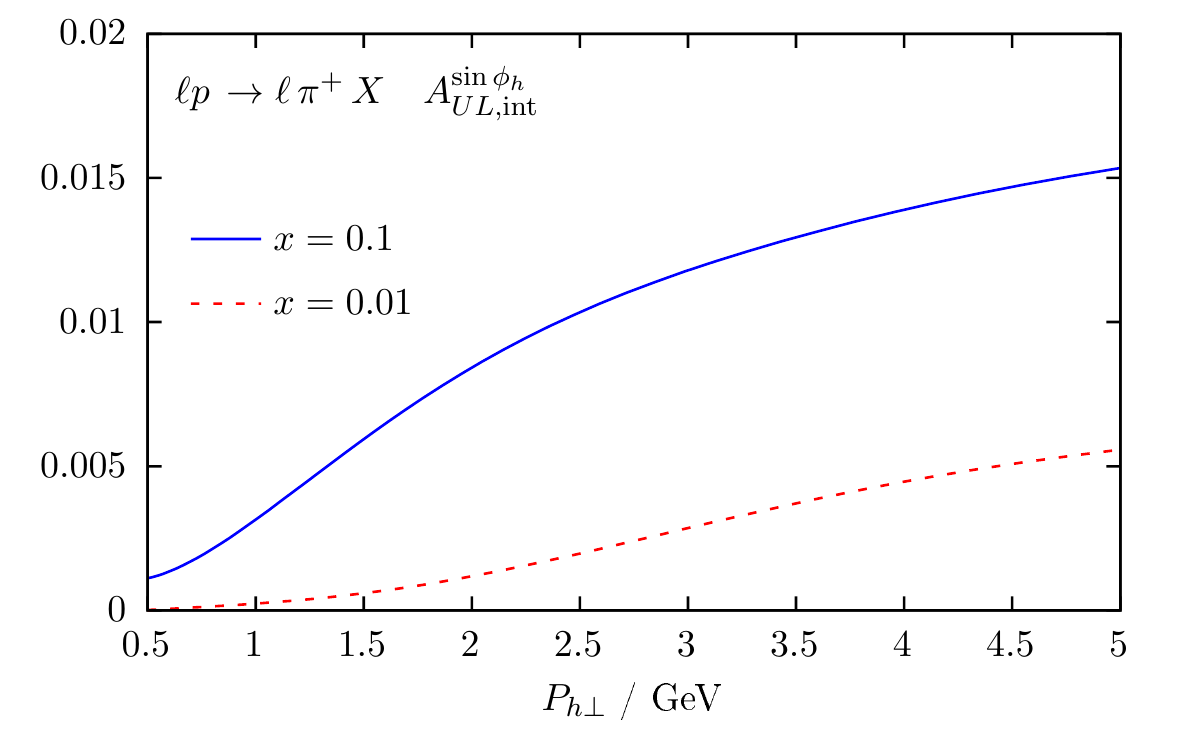}
\includegraphics[scale=0.75]{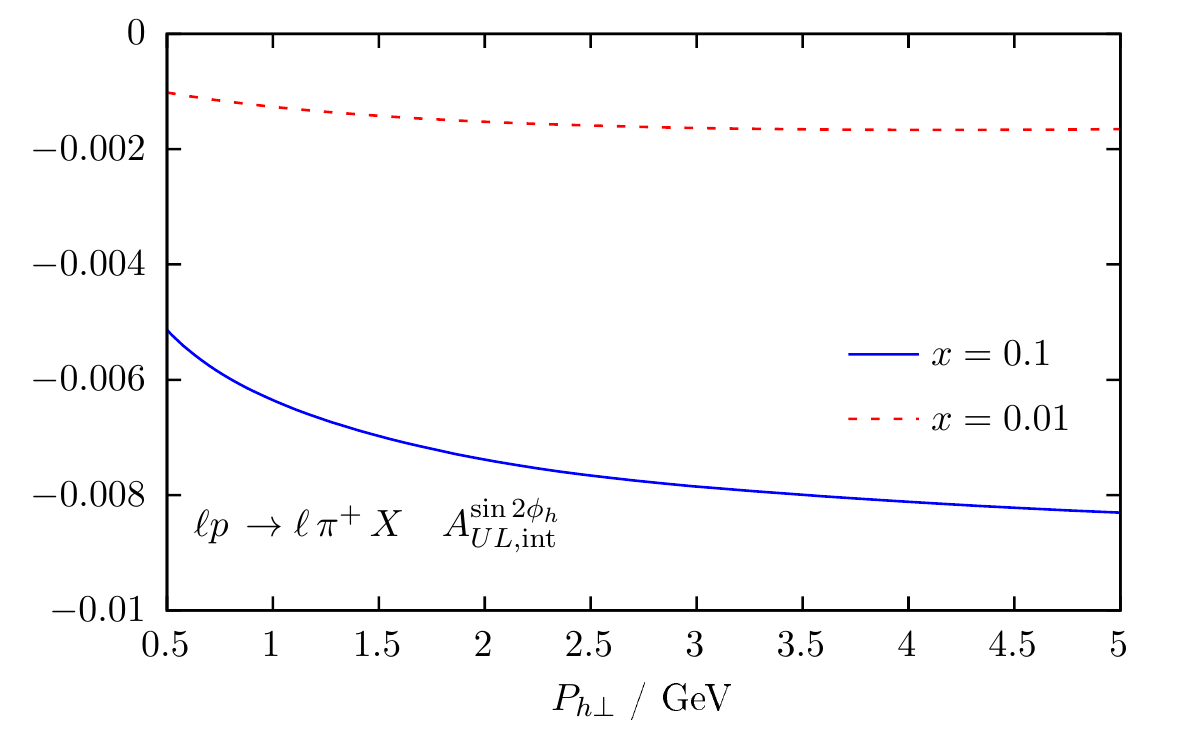}

\includegraphics[scale=0.75]{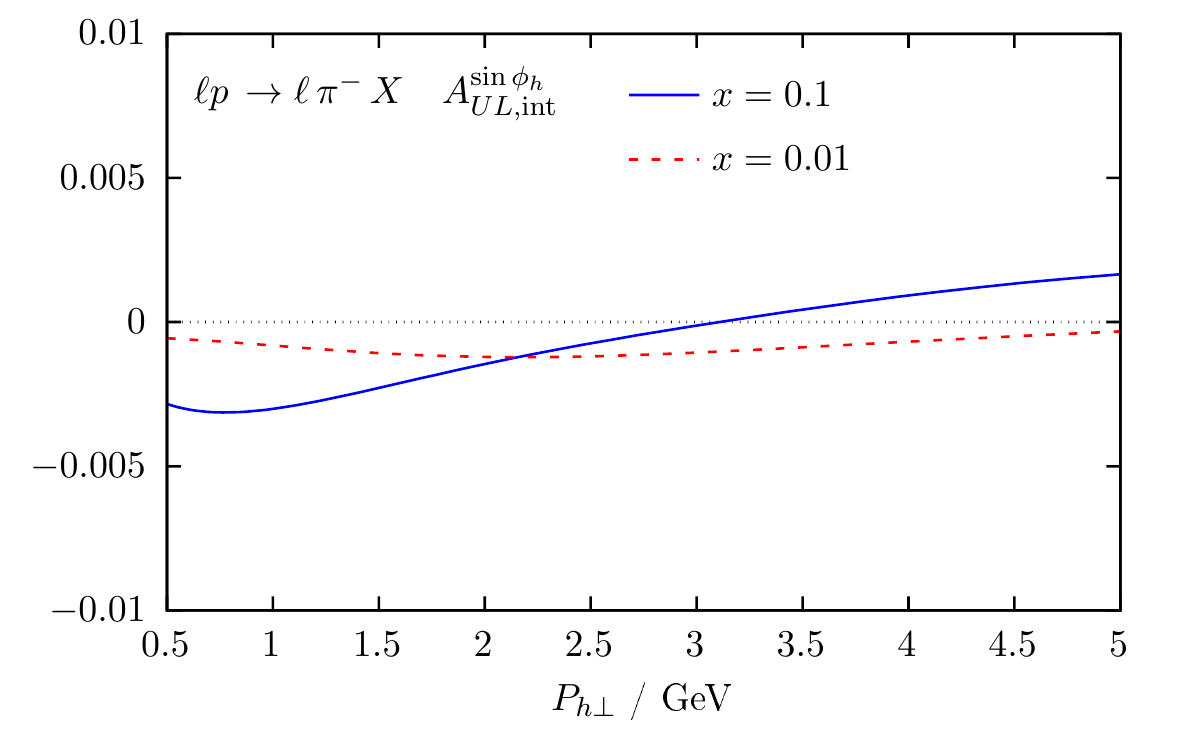}
\includegraphics[scale=0.75]{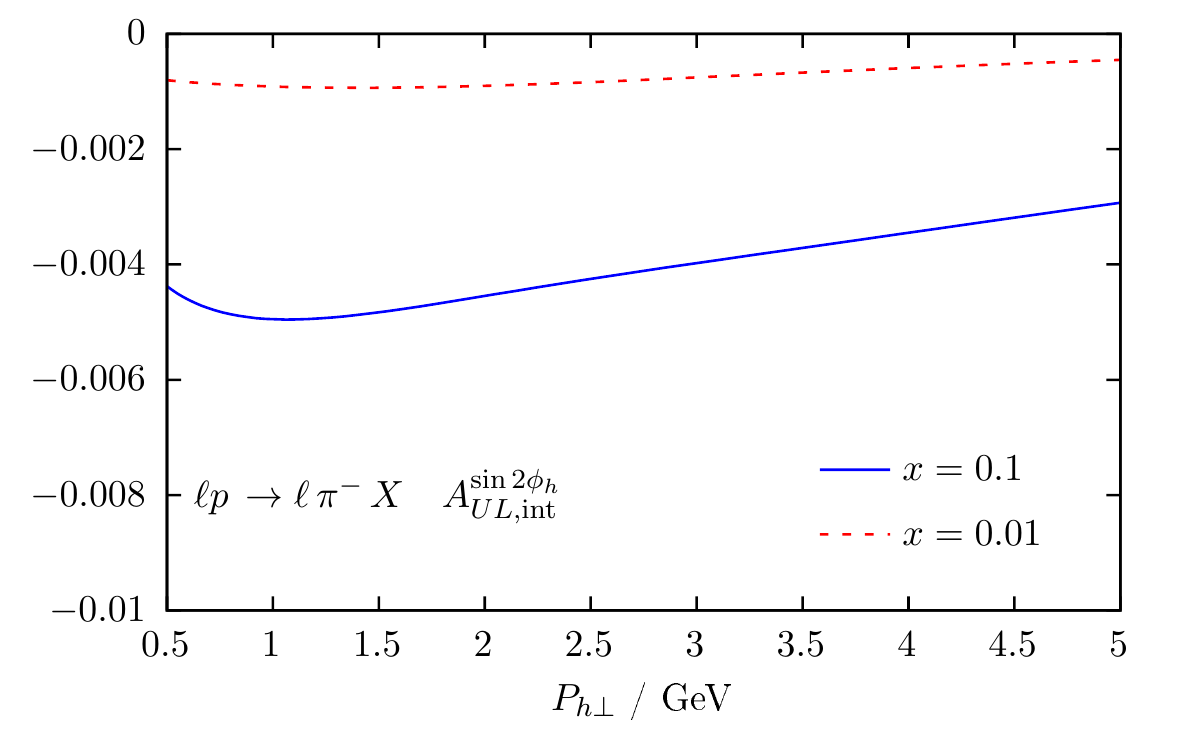}

\caption{\it{T-odd asymmetries as functions of  $P_{h \perp}$ for $x=0.1$ and $x=0.01$ for $\ell p\,\rightarrow \ell\, \pi^{+}\, X$ and $\ell p\,\rightarrow \ell\, \pi^{-}\, X$ 
at the EIC. We have integrated the cross sections 
over $Q^2/(\mathrm{GeV})^2 \in [10, 100]$ for $x=0.1$ (blue solid) and $Q^2/(\mathrm{GeV})^2 \in [2, 10]$ for $x=0.01$ (red dashed), 
and in both cases over $z \in [0.05, 0.8]$.}}
\label{fig:AUL-int} 
\end{figure}

In the following we show results integrated over large bins in $z$ and $Q^{2}$, but differential in $P_{h \perp}$ and $x$.
To this end, we define
\begin{equation}
	A^{\sin n \phi_{h}}_{UL,\mathrm{int}}
		\,\equiv\, \frac{
			\int_{z_{\mathrm{min}}}^{z_{\mathrm{max}}}dz 
			\int_{Q^{2}_{\mathrm{min}}}^{Q^{2}_{\mathrm{max}}}dQ^{2} 
			\int_0^{2\pi} d\phi_{h} \sin(n \phi_{h}) 
			\big[d\sigma^{+}(\phi_{h}) - d\sigma^{-}(\phi_{h})\big]}
		{ \frac{1}{2}\int_{z_{\mathrm{min}}}^{z_{\mathrm{max}}}dz 
			\int_{Q^{2}_{\mathrm{min}}}^{Q^{2}_{\mathrm{max}}}dQ^{2} 
			\int_0^{2\pi}  d\phi_{h} 
			\big[d\sigma^{+}(\phi_{h}) + d\sigma^{-}(\phi_{h})\big]}\quad\quad (n=1,2)\, .
\end{equation}
Figure~\ref{fig:AUL-int} shows $A^{\sin \phi_{h}}_{UL,\mathrm{int}}$ and $A^{\sin 2\phi_{h}}_{UL,\mathrm{int}}$
as functions of $P_{h \perp}$ at the EIC, for fixed values $x=0.1$ (blue solid) and $x=0.01$ (red dashed), for
production of positive and negative pions. As expected for an ${\cal O}(\alpha_s)$ effect, and because of the 
suppression of the polarized parton distributions already mentioned, the asymmetries are quite small, especially
for $x=0.01$. Also, $A^{\sin 2\phi_{h}}_{UL,\mathrm{int}}$ is generally smaller than $A^{\sin \phi_{h}}_{UL,\mathrm{int}}$ 
because of its stronger suppression at low $q_T/Q$ discussed in the previous section.  
We also observe that the asymmetries for positively and negatively charged pions tend to have opposite
signs, which is due to the dominance of the (positive) up-quark helicity parton distribution for $\pi^+$ production,
and of the (negative) down-quark helicity distribution in case of $\pi^-$. 
We note that detailed studies of the uncertainties to be expected for such measurements at the EIC will evidently
require a full analysis that also incorporates efficiencies and detector effects, which is beyond the scope
of our paper. A ballpark estimate provides confidence that even asymmetries of the small size as in Fig.~\ref{fig:AUL-int}
should be resolvable at the EIC. 

\begin{figure}
\includegraphics[scale=0.73]{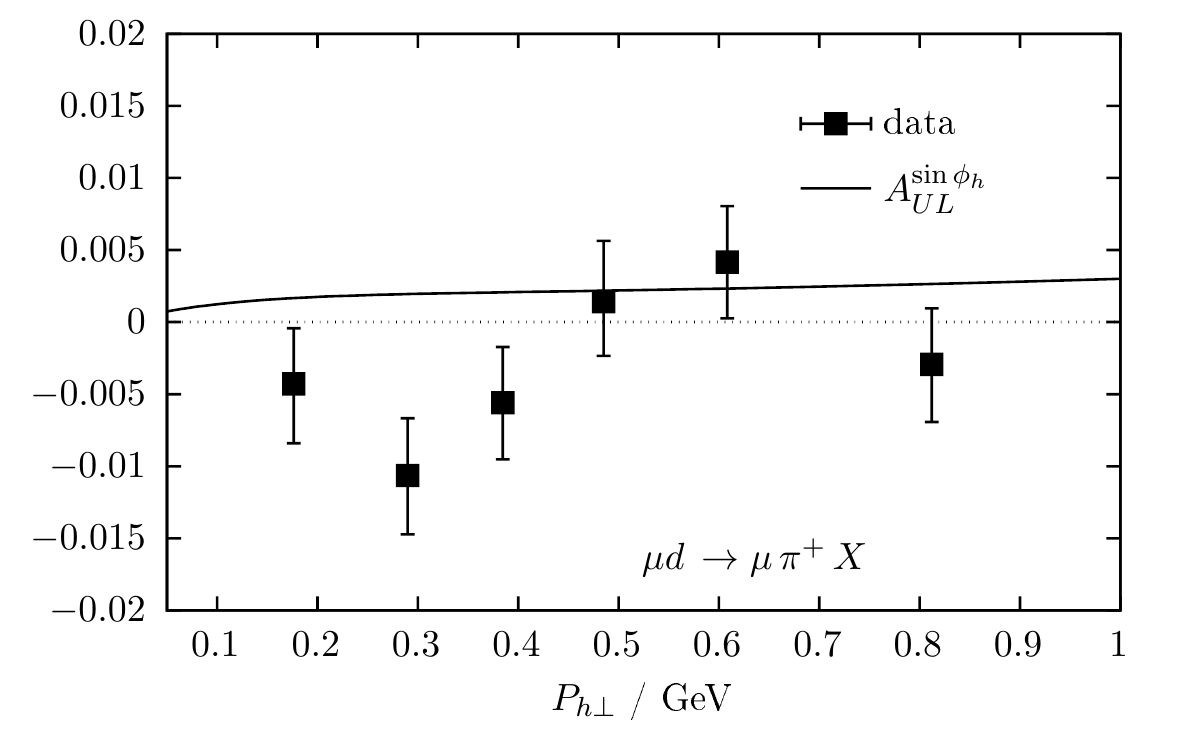}
\includegraphics[scale=0.73]{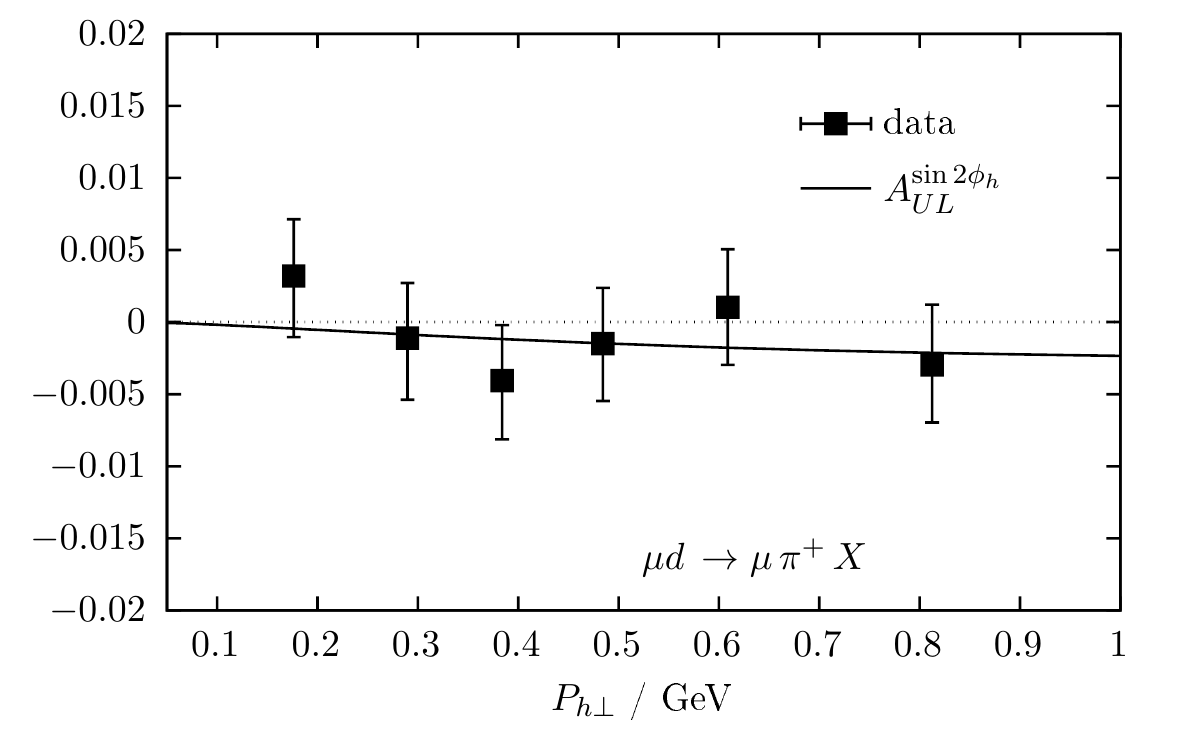}
\caption{\it{Comparison of our numerical estimates for $A^{\sin\phi_{h}}_{UL}$ and $A^{\sin2\phi_{h}}_{UL}$ with data from the COMPASS collaboration \cite{COMPASS:2016klq} in hadron production off longitudinally polarized deuterons at $\sqrt{s} =17.4$ GeV. We show the $P_{h \perp}$ 
distribution of the asymmetry integrated over $x \in [0.004,0.7]$, $z \in [0.01,1]$ and $Q^2/(\mathrm{GeV})^2 \in [1, 100]$.}}
\label{fig:compass-data}
\end{figure}

We proceed by presenting a comparison to data from COMPASS~\cite{COMPASS:2016klq} where the asymmetry 
$A_{UL}$ has been measured in muon scattering off longitudinally polarized deuterons at $\sqrt{s} =17.4$ GeV. We note that COMPASS has considered the production of arbitrary charged hadrons. Sets of fragmentation functions for $h^\pm$ are
not available in DSS14, so we continue to use the pion fragmentation functions. Given that pions dominate the spectrum of produced hadrons
and that fragmentation effects cancel to some extent in the spin asymmetry, this should be more than sufficient for a first comparison. 
We consider the $\pi^{+}$-channel: $\mu d\,\rightarrow \mu\, \pi^{+}\, X$. As in~\cite{COMPASS:2016klq} we integrate over
$x \in [0.004,0.7]$, $z \in [0.01,1]$ and $Q^2/(\mathrm{GeV})^2 \in [1, 100]$ and divide by the value $\vert P_{L} \vert=0.8$ of 
the muon beam polarization. Figure \ref{fig:compass-data} shows the comparisons both for $A^{\sin\phi_{h}}_{UL}$ and for
$A^{\sin2\phi_{h}}_{UL}$. We observe reasonable agreement, given the rather large uncertainties of the data, and keeping
in mind that the $P_{h \perp}$ values are such that arguably a TMD description would appear to be more appropriate. 

\begin{figure}
\vspace*{-4mm}
\centering
\includegraphics[scale=0.53]{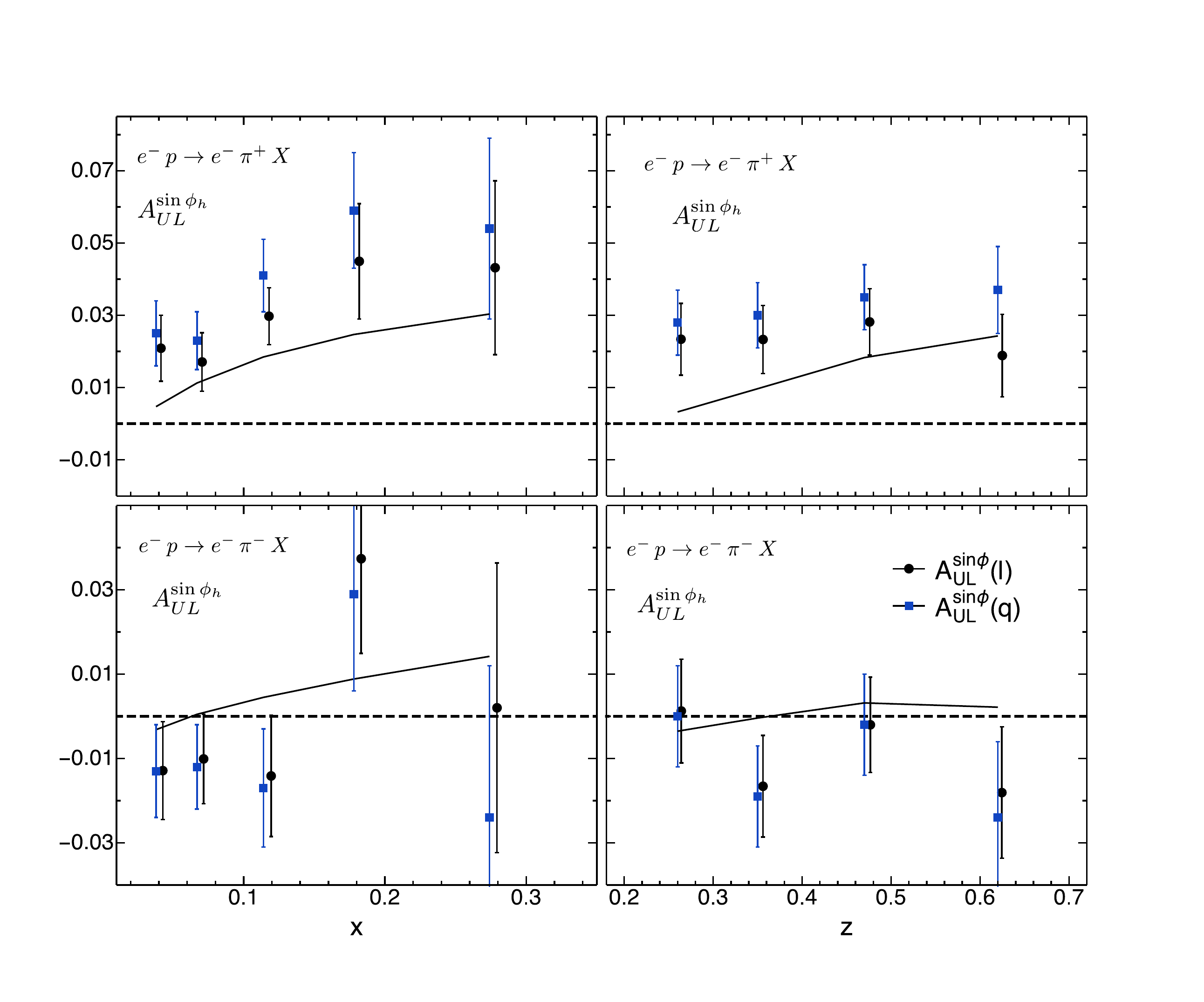}
\vspace*{-5mm}
\caption{\it{Comparison of our calculations of $A^{\sin\phi_{h}}_{UL}$ with data from the HERMES collaboration \cite{HERMES:2005mov} in $\pi^{\pm}$ production off longitudinally polarized protons, as functions of $x$ (left) and $z$ (right).  
$A^{\sin\phi_{h}}_{UL}(l)$ and $A^{\sin\phi_{h}}_{UL}(q)$ represent the asymmetries for target polarization defined relative to the lepton or photon direction, 
respectively.}}
\label{fig:hermes-data}
\end{figure}

We finally also show a comparison to data from the HERMES experiment~\cite{HERMES:1999ryv,HERMES:2005mov} taken
for $\pi^\pm$ production at $\sqrt{s}=7.25$~GeV. 
As we discussed in Sec.~\ref{pSIDIS}, in an actual experiment the target is polarized along (or opposite to) the lepton beam direction. This means
that the measured asymmetry $A_{UL}$ receives contributions from both the longitudinal and transverse spin asymmetries with respect to the direction 
of the virtual photon~\cite{Diehl:2005pc,HERMES:2005mov}, so that $A_{UL}(l)\neq A_{UL}(q)$ (where the arguments $l$ and $q$ denote
target polarization defined relative to the lepton or photon direction, respectively). For HERMES with its relatively modest $Q^2$ values, the 
difference between $A_{UL}(l)$ and $A_{UL}(q)$ -- which is of subleading twist --
is expected to be potentially more pronounced. Combining with data taken with a transversely 
polarized target, HERMES has in fact been able to provide an extraction of $A_{UL}(q)$~\cite{HERMES:2005mov}. Figure~\ref{fig:hermes-data} shows
both sets of HERMES data, $A_{UL}^{\sin\phi_{h}}(l)$ and $A_{UL}^{\sin\phi_{h}}(q)$, compared to our calculations of $A_{UL}^{\sin\phi_{h}}(q)$. 
We show the comparisons as functions of $x$ and $z$, using the mean values of $x$, $z$, $Q^2$ and $P_{h\perp}$ for each point reported in Table~1 
of \cite{HERMES:2005mov}. One can see that for positively charged pions the differences between $A_{UL}^{\sin\phi_{h}}(l)$ and $A_{UL}^{\sin\phi_{h}}(q)$ 
are quite large, while they are small for $\pi^-$ production. We also notice that our calculations reproduce the trend of the data rather 
well overall, despite the fact that HERMES accessed only rather small transverse momenta. 

\begin{figure}
\centering
\includegraphics[scale=0.53]{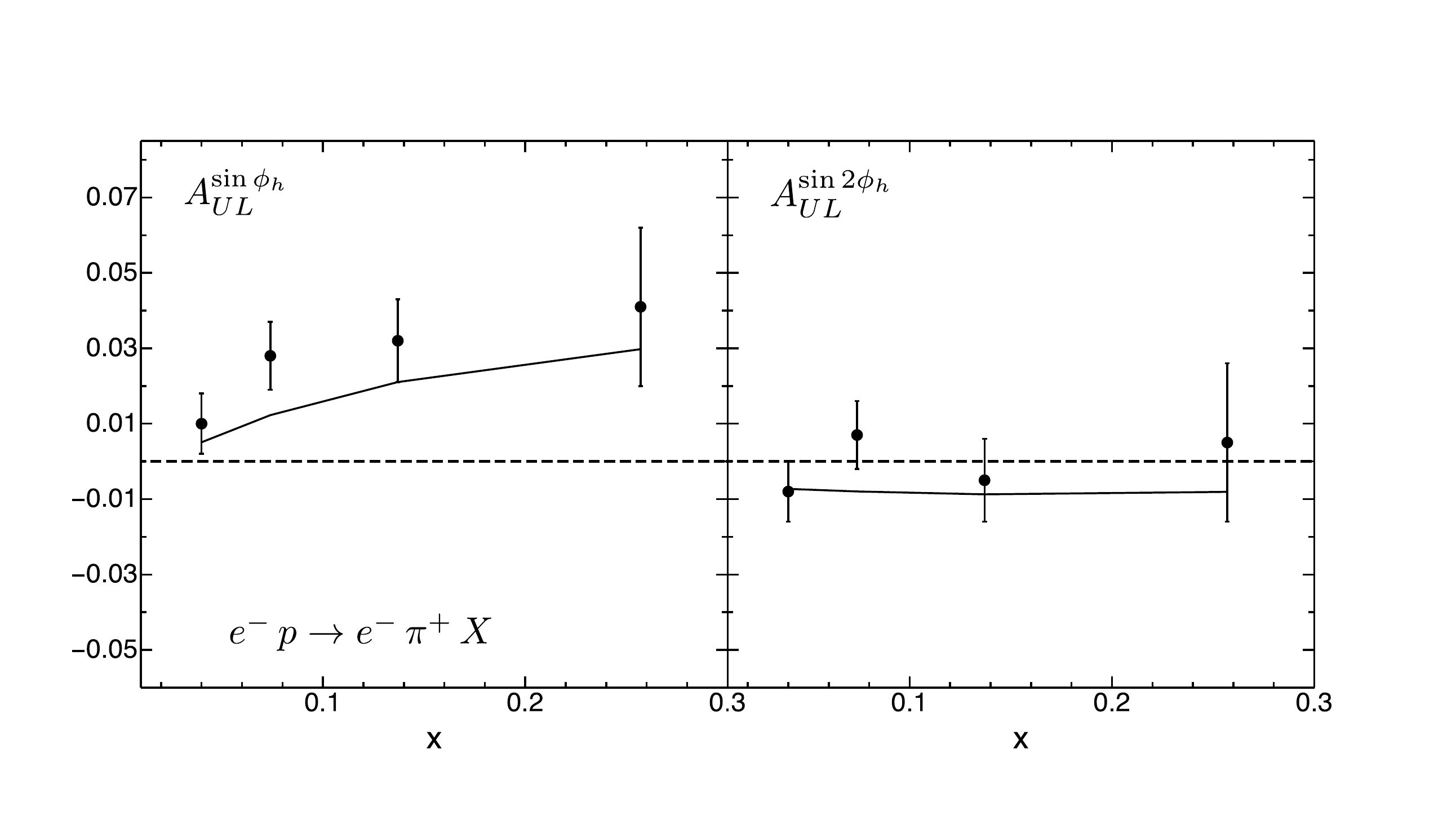}
\vspace*{-5mm}
\caption{\it{Numerical results for $A^{\sin\phi_{h}}_{UL}$ and $A^{\sin2\phi_{h}}_{UL}$ compared to HERMES data~\cite{HERMES:1999ryv} for 
$\pi^{+}$ production off longitudinally polarized protons. The data have not been corrected for the polarization direction.}}
\label{fig:old-hermes-data}
\end{figure}

HERMES data for the $\sin2\phi_{h}$ asymmetry are available only without correction for the polarization direction~\cite{HERMES:1999ryv}. 
The right part of Fig.~\ref{fig:old-hermes-data} shows the data for $A^{\sin2\phi_{h}}_{UL}(l)$ for $\pi^+$ production, compared to our 
calculations. Here we have used the mean values of $x$ and $Q^{2}$ for each point as given in~\cite{HERMES:1999ryv}, 
while adopting $\braket{z}$ and $\braket{P_{h\perp}}$ from~\cite{HERMES:2005mov}. To facilitate comparison with the $\sin\phi_{h}$ asymmetry,
we show the corresponding results for  $A^{\sin\phi_{h}}_{UL}$ on the left side of Fig.~\ref{fig:old-hermes-data}. As before, the $\sin2\phi_{h}$ asymmetry
is smaller. Our calculations are overall in fair agreement with the data for both asymmetries; in particular, they nicely capture the 
difference in magnitude between the $\sin\phi_{h}$ and $\sin2\phi_{h}$ components. The situation thus appears to be different from that 
for the $\cos 2\phi_{h}$ unpolarized structure function $F^{\cos2\phi_{h}}_{UU}$ analyzed in \cite{Barone:2008tn}, for which 
the perturbative $\mathcal{O}(\alpha_{s})$ prediction for HERMES kinematics was shown to be negligible compared to higher-twist effects.
It should be stressed again, however, that the data shown in Fig.~\ref{fig:old-hermes-data} have not been corrected for the polarization direction,
and according to Fig.~\ref{fig:hermes-data} such correction effects are expected to be particularly important in the case of $\pi^+$ production.

\section{Conclusions \label{conc}}

We have presented a perturbative calculation for the single-spin asymmetry $A_{UL}$ in semi-inclusive deep-inelastic scattering,
which may be measured by scattering unpolarized leptons off longitudinally polarized nucleons. This asymmetry is interesting because, 
in the absence of parity violation, it is T-odd and receives perturbative contributions only via QCD loop effects.  
Also, it is sensitive to the proton's helicity parton distributions, despite the fact that it is measured with an
unpolarized lepton beam. 
Our calculation builds on the large body of previous work on T-odd effects in hard scattering, opening a new
avenue for future measurements at the EIC.

We have provided compact expressions for the T-odd contributions in the various partonic channels. We have used
these to derive the low-transverse-momentum behavior of the T-odd terms, which shows striking features. 
Our results add new information on the relations between TMDs and perturbative hard scattering which had been missing so far. 
They may be used for comparisons to detailed quantitative predictions to be obtained in the future within the TMD formalism. 

Our phenomenological calculations reveal the expected relatively small size of the T-odd asymmetries.
We have made predictions for the asymmetries at the future electron-ion-collider, where it should
be possible to explore them.  
Our results are also broadly consistent with available COMPASS and HERMES data, although the
applicability of a purely hard-scattering picture is questionable here. All in all we hope
that our paper will contribute to the long-standing quest to establish and understand T-odd
effects in QCD.

\section*{Acknowledgments}

We thank A.~Bacchetta, M.~Diehl, M.~Schlegel and A.~Vladimirov for useful discussions. 
A.S. is grateful to the University of the Basque Country, Bilbao, for hospitality.
This study was supported in part by Deutsche Forschungsgemeinschaft (DFG) through the Research Unit FOR 2926.

\appendix
\section{Appendix: Perturbative results for the beam-spin asymmetry $A_{LU}$ \label{APP}}

The cross section for the beam-spin asymmetry $A_{LU}$ may be written as~\cite{Bacchetta:2006tn}:
\begin{eqnarray}
\label{spindepcrossLU}
\frac{d^{\,5}\Delta\sigma^h}{dx\; dy\; dz\; dP^{2}_{h\perp}\, d\phi_{h}}&=&
\frac{1}{2}\left(\frac{d^{\,5}\sigma^h_+}{dx\; dy\; dz\; dP^{2}_{h\perp}\,d\phi_{h}} - 
\frac{d^{\,5}\sigma^h_-}{dx\; dy\; dz\; dP^{2}_{h\perp}\,d\phi_{h}} \right)\nn\\[2mm]
	&=&\frac{\pi \alpha^{2}y}{x Q^{2}}\,
	\sqrt{\frac{2\varepsilon}{1- \varepsilon}}\,F^{\,\sin\phi_{h}}_{LU}\sin(\phi_{h})\,.
\end{eqnarray}
As is well-known, there is no term with $\sin(2\phi_h)$ for single-photon exchange. 
Writing the structure function $F^{\,\sin\phi_{h}}_{LU}$ as
\beeq\label{structfunc1}
F^{\,\sin\phi_{h}}_{LU}&=&\left(\frac{ \alpha_{s}(\mu^2)}{2\pi}\right)^2\,\frac{x}{Q^{2}z^2}
\sum_{\begin{subarray}{c}
a,b \\
=\,q,\bar{q},g \end{subarray}} 
\int_{x}^{1} \frac{d\hat{x}}{\hat{x}}
	\int_{z}^{1} \frac{d\hat{z}}{\hat{z}}\,\Delta f_{a}\left(\frac{x}{\hat{x}},\mu^{2}\right)\,
	 		C^{\,\sin\phi_h\,,a\to b}_{LU}(\hat{x}, \hat{z})
		\,D^h_{b}\left(\frac{z}{\hat{z}},\mu^{2}\right)\nn\\[2mm]
		&\times&\;\delta\left( \frac{q_{T}^{2}}{Q^{2}} - \frac{(1 - \hat{x})(1 - \hat{z})}{\hat{x} \hat{z}}\right)\,,
\eeeq
we have from Refs.~\cite{Hagiwara:1982cq,Korner:2000zr}:
\begin{align}
	C^{\,\sin\phi_h,q \to q}_{LU}(\hat{x}, \hat{z})
		&= -e_q^2 \,C_{F}
			\left( C_{A}(1-\hat{x}) - C_{F}(1-\hat{x}+\hat{z} +\hat{x}\hat{z})+ (C_{A}-2C_{F})\frac{\hat{z}\ln\hat{z}}{1-\hat{z}} 
			\right) \frac{Q}{q_T}\, ,\nn\\[2mm]
C^{\,\sin\phi_h,q\to  g}_{LU}(\hat{x}, \hat{z})
		&= e_q^2 \,C_F \frac{(1-\hat{z})}{\hat{z}} \biggl( C_A (1-\hat{x})+C_F (\hat{x} \hat{z}+\hat{z}-2)  \nn\\[2mm]
		&\left.	+ (C_A - 2 C_F)\frac{ (1-\hat{z})  \ln (1-\hat{z})}{\hat{z}}\right)\frac{Q}{q_T} \, ,\nn\\[2mm]
C^{\,\sin\phi_h, g\rightarrow q }_{LU}(\hat{x}, \hat{z})
		&= -e_q^2\,(C_A-2 C_F)\frac{1-\hat{x}}{2\hat{z}^2}\biggl(-\hat{x} \hat{z} (1-2 \hat{z})-(1-\hat{x}) \ln (1-\hat{z})\nn\\[2mm]
		&\left.	+(1-\hat{x})\frac{ \hat{z} \ln (\hat{z})}{1-\hat{z}} \right)\frac{Q}{q_T}\, .
		\label{CLUphi}
\end{align}
Carrying out the expansions for low $q_T/Q$ we find, up to corrections of order $q_T/Q$:
\begin{align}\label{smallqt0}
&\hspace*{-1mm}\big(f_q\otimes C^{\,\sin\phi_h, q\rightarrow q}_{LU} \otimes D_q^h \big)(x,z)\nn \\[2mm]
	&=\,e_q^2\,\frac{Q}{q_{T}}\,\frac{C_A}{2}\left\{
	-C_F\left( 2 \,\ln\left( \frac{q_T^{2}}{Q^{2}}\right)+3\right) 
	 f_{q}\left(x \right)	D^h_{q}\left(z \right) \right.\nn\\[2mm]
	&+ \,D^h_{q}\left(z\right)\int_{x}^{1}d\hat{x}\;
			\delta P_{qq}(\hat{x})\,
			 f_{q}\left(\frac{x}{\hat{x}} \right)+\left. f_{q}\left(x\right) 
			\int_{z}^{1} d\hat{z}\;\delta P_{qq}(\hat{z})\,D^h_{q}\left(\frac{z}{\hat{z}} \right)
			\right\}\nn\\[2mm]
			&-\,e_q^2\,\frac{Q}{q_{T}}\frac{C_F}{C_A}\,
			 f_{q}\left(x\right) 
			\int_{z}^{1} d\hat{z}\,
			\frac{\hat{z}}{1-\hat{z}}\,
			\left( 1 + \frac{\ln \hat{z}}{1-\hat{z}}\right)
			D^h_{q}\left(\frac{z}{\hat{z}} \right)\,,\nn\\[2mm]
			&\hspace*{-1mm}\big(f_q\otimes C^{\,\sin\phi_h, q\rightarrow g}_{LU} \otimes D_g^h \big)(x,z)\nn \\[2mm]
	&=e_q^2\,\frac{Q}{q_{T}}C_F\,f_{q}\left(x\right) 
			\int_{z}^{1} d\hat{z}\,
			\frac{1-\hat{z}}{\hat{z}^2}\,
			\big((C_A-2 C_F) \ln(1-\hat{z})-2 C_F \hat{z}\big)
			D_g^h\left(\frac{z}{\hat{z}} \right)\,,\nn\\[2mm]
&\hspace*{-1mm}\big(f_g\otimes C^{\,\sin\phi_h, g\rightarrow q}_{LU} \otimes D_q^h \big)(x,z)]\nn \\[2mm]
	&=-\frac{e_q^2}{2}\,\frac{Q}{q_{T}}
	(C_A-2 C_{F})D^h_{q}\left(z\right)
	\int_{x}^{1}d\hat{x}\;
			f_{g}\left(\frac{x}{\hat{x}} \right)\left[ (1-\hat{x})\ln\left(\frac{Q^2}{q_T^{2}}\right)+ (1-\hat{x}) \ln\left(\frac{1-\hat{x}}{\hat{x}}\right) +2x-1  \right]\,,
\end{align}
where the convolution has been defined in~(\ref{dconv}) and the transversity splitting function in~(\ref{transv}).

\newpage


\begin{thebibliography}{99}

\bibitem{GrossePerdekamp:2015xdx}
{\it For review, see:} M.~Grosse Perdekamp and F.~Yuan,
Ann. Rev. Nucl. Part. Sci. \textbf{65}, 429 (2015)
[arXiv:1510.06783 [hep-ph]].

\bibitem{Sivers:1989cc}
D.~W.~Sivers,
Phys. Rev. D \textbf{41}, 83 (1990).

\bibitem{Brodsky:2002cx}
S.~J.~Brodsky, D.~S.~Hwang and I.~Schmidt,
Phys. Lett. B \textbf{530}, 99 (2002)
[arXiv:hep-ph/0201296 [hep-ph]].

\bibitem{Collins:2002kn}
J.~C.~Collins,
Phys. Lett. B \textbf{536}, 43 (2002)
[arXiv:hep-ph/0204004 [hep-ph]].

\bibitem{DeRujula:1971nnp}
A.~De Rujula, J.~M.~Kaplan and E.~De Rafael,
Nucl. Phys. B \textbf{35}, 365 (1971).

\bibitem{DeRujula:1978bz}
A.~De Rujula, R.~Petronzio and B.~E.~Lautrup,
Nucl. Phys. B \textbf{146}, 50 (1978).

\bibitem{Fabricius:1980wg}
K.~Fabricius, I.~Schmitt, G.~Kramer and G.~Schierholz,
Phys. Rev. Lett. \textbf{45}, 867 (1980).

\bibitem{Korner:1980np}
J.~G.~K\"{o}rner, G.~Kramer, G.~Schierholz, K.~Fabricius and I.~Schmitt,
Phys. Lett. B \textbf{94}, 207 (1980).

\bibitem{Hagiwara:1981qn}
K.~Hagiwara, K.~i.~Hikasa and N.~Kai,
Phys. Rev. Lett. \textbf{47}, 983 (1981).

\bibitem{Hagiwara:1982cq}
K.~Hagiwara, K.~i.~Hikasa and N.~Kai,
Phys. Rev. D \textbf{27}, 84 (1983).

\bibitem{Pire:1983tv}
B.~Pire and J.~P.~Ralston,
Phys. Rev. D \textbf{28}, 260 (1983).

\bibitem{Hagiwara:1984hi}
K.~Hagiwara, K.~i.~Hikasa and N.~Kai,
Phys. Rev. Lett. \textbf{52}, 1076 (1984).

\bibitem{Bilal:1990wi}
A.~Bilal, E.~Masso and A.~De Rujula,
Nucl. Phys. B \textbf{355}, 549 (1991).

\bibitem{Carlitz:1992fv}
R.~D.~Carlitz and R.~S.~Willey,
Phys. Rev. D \textbf{45}, 2323 (1992).

\bibitem{Brandenburg:1995nv}
A.~Brandenburg, L.~J.~Dixon and Y.~Shadmi,
Phys. Rev. D \textbf{53}, 1264 (1996)
[arXiv:hep-ph/9505355 [hep-ph]].

\bibitem{Ahmed:1999ix}
M.~Ahmed and T.~Gehrmann,
Phys. Lett. B \textbf{465}, 297 (1999)
[arXiv:hep-ph/9906503 [hep-ph]].

\bibitem{Korner:2000zr}
J.~G.~K\"{o}rner, B.~Melic and Z.~Merebashvili,
Phys. Rev. D \textbf{62}, 096011 (2000)
[arXiv:hep-ph/0002302 [hep-ph]].

\bibitem{Hagiwara:2007sz}
K.~Hagiwara, K.~Mawatari and H.~Yokoya,
JHEP \textbf{12}, 041 (2007)
[arXiv:0707.3194 [hep-ph]].

\bibitem{Frederix:2014cba}
R.~Frederix, K.~Hagiwara, T.~Yamada and H.~Yokoya,
Phys. Rev. Lett. \textbf{113}, 152001 (2014)
[arXiv:1407.1016 [hep-ph]].

\bibitem{Benic:2019zvg}
S.~Benic, Y.~Hatta, H.~n.~Li and D.~J.~Yang,
Phys. Rev. D \textbf{100}, 094027 (2019)
[arXiv:1909.10684 [hep-ph]].

\bibitem{aicher}
M.~Aicher, Phd thesis, 2011, Universit\"at Regensburg.

\bibitem{abele}
M.~Abele, MSc thesis, 2019, University of T\"{u}bingen.

\bibitem{Piacenza:2020sst}
F.~Piacenza, Phd thesis, 2020, Universit\`{a} di Pavia.

\bibitem{Qiu:1998ia}
J.~w.~Qiu and G.~F.~Sterman,
Phys. Rev. D \textbf{59}, 014004 (1999)
[arXiv:hep-ph/9806356 [hep-ph]].

\bibitem{HERMES:1999ryv}
A.~Airapetian \textit{et al.} [HERMES],
Phys. Rev. Lett. \textbf{84}, 4047 (2000)
[arXiv:hep-ex/9910062 [hep-ex]].

\bibitem{HERMES:2001hbj}
A.~Airapetian \textit{et al.} [HERMES],
Phys. Rev. D \textbf{64}, 097101 (2001)
[arXiv:hep-ex/0104005 [hep-ex]].

\bibitem{HERMES:2002buj}
A.~Airapetian \textit{et al.} [HERMES],
Phys. Lett. B \textbf{562}, 182 (2003)
[arXiv:hep-ex/0212039 [hep-ex]].

\bibitem{HERMES:2005mov}
A.~Airapetian \textit{et al.} [HERMES],
Phys. Lett. B \textbf{622}, 14 (2005)
[arXiv:hep-ex/0505042 [hep-ex]].

\bibitem{CLAS:2010fns}
H.~Avakian \textit{et al.} [CLAS],
Phys. Rev. Lett. \textbf{105}, 262002 (2010)
[arXiv:1003.4549 [hep-ex]].

\bibitem{COMPASS:2016klq}
C.~Adolph \textit{et al.} [COMPASS],
Eur. Phys. J. C \textbf{78}, 952 (2018)
[erratum: Eur. Phys. J. C \textbf{80}, 298 (2020)]
[arXiv:1609.06062 [hep-ex]].

\bibitem{Savin:2016gah}
I.~A.~Savin [COMPASS],
J. Phys. Conf. Ser. \textbf{678}, 012063 (2016)

\bibitem{COMPASS:2010nak}
M.~G.~Alekseev \textit{et al.} [COMPASS],
Eur. Phys. J. C \textbf{70}, 39 (2010)
[arXiv:1007.1562 [hep-ex]].

\bibitem{Aschenauer:2019kzf}
E.~C.~Aschenauer, I.~Borsa, R.~Sassot and C.~Van Hulse,
Phys. Rev. D \textbf{99}, 094004 (2019)
doi:10.1103/PhysRevD.99.094004
[arXiv:1902.10663 [hep-ph]].

\bibitem{Rogers:2015sqa}
T.~C.~Rogers,
Eur. Phys. J. A \textbf{52}, 153 (2016)
[arXiv:1509.04766 [hep-ph]].

\bibitem{Angeles-Martinez:2015sea}
R.~Angeles-Martinez, A.~Bacchetta, I.~I.~Balitsky, D.~Boer, M.~Boglione, R.~Boussarie, F.~A.~Ceccopieri, I.~O.~Cherednikov, P.~Connor and M.~G.~Echevarria, \textit{et al.}
Acta Phys. Polon. B \textbf{46}, 2501 (2015)
[arXiv:1507.05267 [hep-ph]].

\bibitem{Bacchetta:2004zf}
A.~Bacchetta, P.~J.~Mulders and F.~Pijlman,
Phys. Lett. B \textbf{595}, 309-317 (2004)
[arXiv:hep-ph/0405154 [hep-ph]].

\bibitem{Bacchetta:2006tn}
A.~Bacchetta, M.~Diehl, K.~Goeke, A.~Metz, P.~J.~Mulders and M.~Schlegel,
JHEP \textbf{02}, 093 (2007)
[arXiv:hep-ph/0611265 [hep-ph]].

\bibitem{Boglione:2000jk}
M.~Boglione and P.~J.~Mulders,
Phys. Lett. B \textbf{478}, 114 (2000)
[arXiv:hep-ph/0001196 [hep-ph]].

\bibitem{Ma:2000ip}
B.~Q.~Ma, I.~Schmidt and J.~J.~Yang,
Phys. Rev. D \textbf{63}, 037501 (2001)
[arXiv:hep-ph/0009297 [hep-ph]].

\bibitem{Ma:2001ie}
B.~Q.~Ma, I.~Schmidt and J.~J.~Yang,
Phys. Rev. D \textbf{65}, 034010 (2002)
[arXiv:hep-ph/0110324 [hep-ph]].

\bibitem{Efremov:2003tf}
A.~V.~Efremov, K.~Goeke and P.~Schweitzer,
Phys. Lett. B \textbf{568}, 63 (2003)
[arXiv:hep-ph/0303062 [hep-ph]].

\bibitem{Boffi:2009sh}
S.~Boffi, A.~V.~Efremov, B.~Pasquini and P.~Schweitzer,
Phys. Rev. D \textbf{79}, 094012 (2009)
[arXiv:0903.1271 [hep-ph]].

\bibitem{Zhu:2011zza}
J.~Zhu and B.~Q.~Ma,
Phys. Lett. B \textbf{696}, 246 (2011)
[arXiv:1104.4564 [hep-ph]].

\bibitem{Lu:2011pt}
Z.~Lu, B.~Q.~Ma and J.~She,
Phys. Rev. D \textbf{84}, 034010 (2011)
[arXiv:1104.5410 [hep-ph]].

\bibitem{Li:2021mmi}
H.~Li and Z.~Lu,
[arXiv:2111.03840 [hep-ph]].

\bibitem{Bacchetta:2008xw}
A.~Bacchetta, D.~Boer, M.~Diehl and P.~J.~Mulders,
JHEP \textbf{08}, 023 (2008)
[arXiv:0803.0227 [hep-ph]].

\bibitem{Yuan:2003gu}
F.~Yuan,
Phys. Lett. B \textbf{589}, 28-34 (2004)
[arXiv:hep-ph/0310279 [hep-ph]].

\bibitem{Afanasev:2006gw}
A.~V.~Afanasev and C.~E.~Carlson,
Phys. Rev. D \textbf{74}, 114027 (2006)
[arXiv:hep-ph/0603269 [hep-ph]].

\bibitem{Ji:2006ub}
X.~Ji, J.~W.~Qiu, W.~Vogelsang and F.~Yuan,
Phys. Rev. Lett. \textbf{97}, 082002 (2006)
[arXiv:hep-ph/0602239 [hep-ph]].

\bibitem{Ji:2006br}
X.~Ji, J.~W.~Qiu, W.~Vogelsang and F.~Yuan,
Phys. Lett. B \textbf{638}, 178-186 (2006)
[arXiv:hep-ph/0604128 [hep-ph]].

\bibitem{Yuan:2009dw}
F.~Yuan and J.~Zhou,
Phys. Rev. Lett. \textbf{103}, 052001 (2009)
[arXiv:0903.4680 [hep-ph]].

\bibitem{Kanazawa:2015ajw}
K.~Kanazawa, Y.~Koike, A.~Metz, D.~Pitonyak and M.~Schlegel,
Phys. Rev. D \textbf{93}, 054024 (2016)
[arXiv:1512.07233 [hep-ph]].

\bibitem{Moos:2020wvd}
V.~Moos and A.~Vladimirov,
JHEP \textbf{12}, 145 (2020)
[arXiv:2008.01744 [hep-ph]].

\bibitem{Scimemi:2019gge}
I.~Scimemi, A.~Tarasov and A.~Vladimirov,
JHEP \textbf{05}, 125 (2019)
[arXiv:1901.04519 [hep-ph]].

\bibitem{Vladimirov:2021hdn}
A.~Vladimirov, V.~Moos and I.~Scimemi,
JHEP \textbf{01}, 110 (2022)
[arXiv:2109.09771 [hep-ph]].

\bibitem{Scimemi:2018mmi}
I.~Scimemi and A.~Vladimirov,
Eur. Phys. J. C \textbf{78}, 802 (2018)
[arXiv:1804.08148 [hep-ph]].

\bibitem{Ebert:2021jhy}
M.~A.~Ebert, A.~Gao and I.~W.~Stewart,
[arXiv:2112.07680 [hep-ph]].

\bibitem{Rodini:2022wki}
S.~Rodini and A.~Vladimirov,
[arXiv:2204.03856 [hep-ph]].

\bibitem{Diehl:2005pc}
M.~Diehl and S.~Sapeta,
Eur. Phys. J. C \textbf{41}, 515 (2005)
[arXiv:hep-ph/0503023 [hep-ph]].

\bibitem{Mendez:1978zx}
A.~Mendez,
Nucl. Phys. B \textbf{145}, 199 (1978).

\bibitem{Karpman:1968gvz}
G.~Karpman, R.~Leonardi and F.~Strocchi,
Phys. Rev. \textbf{174}, 1957 (1968)

\bibitem{Cannata:1970br}
F.~Cannata, R.~Leonardi and F.~Strocchi,
Phys. Rev. D \textbf{1}, 191 (1970)

\bibitem{tHooft:1972tcz}
G.~'t Hooft and M.~J.~G.~Veltman,
Nucl. Phys. B \textbf{44}, 189 (1972).

\bibitem{Breitenlohner:1977hr}
P.~Breitenlohner and D.~Maison,
Commun. Math. Phys. \textbf{52}, 11 (1977).

\bibitem{Jamin:1991dp}
M.~Jamin and M.~E.~Lautenbacher,
Comput. Phys. Commun. \textbf{74}, 265 (1993).

\bibitem{Patel:2016fam}
H.~H.~Patel,
Comput. Phys. Commun. \textbf{218}, 66 (2017)
[arXiv:1612.00009 [hep-ph]].

\bibitem{Meng:1991da}
R.~b.~Meng, F.~I.~Olness and D.~E.~Soper,
Nucl. Phys. B \textbf{371}, 79 (1992).

\bibitem{Meng:1995yn}
R.~Meng, F.~I.~Olness and D.~E.~Soper,
Phys. Rev. D \textbf{54}, 1919 (1996)
[arXiv:hep-ph/9511311 [hep-ph]].

\bibitem{Boer:2006eq}
D.~Boer and W.~Vogelsang,
Phys. Rev. D \textbf{74}, 014004 (2006)
[arXiv:hep-ph/0604177 [hep-ph]].

\bibitem{Daleo:2004pn}
A.~Daleo, D.~de Florian and R.~Sassot,
Phys. Rev. D \textbf{71}, 034013 (2005)
[arXiv:hep-ph/0411212 [hep-ph]].

\bibitem{Kniehl:2004hf}
B.~A.~Kniehl, G.~Kramer and M.~Maniatis,
Nucl. Phys. B \textbf{711}, 345 (2005)
[erratum: Nucl. Phys. B \textbf{720}, 231 (2005)]
[arXiv:hep-ph/0411300 [hep-ph]].

\bibitem{Wang:2019bvb}
B.~Wang, J.~O.~Gonzalez-Hernandez, T.~C.~Rogers and N.~Sato,
Phys. Rev. D \textbf{99}, 094029 (2019)
[arXiv:1903.01529 [hep-ph]].

\bibitem{deFlorian:2008mr}
D.~de Florian, R.~Sassot, M.~Stratmann and W.~Vogelsang,
Phys. Rev. Lett. \textbf{101}, 072001 (2008)
[arXiv:0804.0422 [hep-ph]].

\bibitem{deFlorian:2009vb}
D.~de Florian, R.~Sassot, M.~Stratmann and W.~Vogelsang,
Phys. Rev. D \textbf{80}, 034030 (2009)
[arXiv:0904.3821 [hep-ph]].

\bibitem{deFlorian:2014xna}
D.~de Florian, R.~Sassot, M.~Epele, R.~J.~Hern\'andez-Pinto and M.~Stratmann,
Phys. Rev. D \textbf{91}, 014035 (2015)
[arXiv:1410.6027 [hep-ph]].

\bibitem{NNPDF:2017mvq}
R.~D.~Ball \textit{et al.} [NNPDF],
Eur. Phys. J. C \textbf{77}, 663 (2017)
[arXiv:1706.00428 [hep-ph]].

\bibitem{Buckley:2014ana}
A.~Buckley, J.~Ferrando, S.~Lloyd, K.~Nordstr\"om, B.~Page, M.~R\"ufenacht, M.~Sch\"onherr and G.~Watt,
Eur. Phys. J. C \textbf{75}, 132 (2015)
[arXiv:1412.7420 [hep-ph]].

\bibitem{Barone:2008tn}
V.~Barone, A.~Prokudin and B.~Q.~Ma,
Phys. Rev. D \textbf{78}, 045022 (2008)
[arXiv:0804.3024 [hep-ph]].
 
\end{thebibliography}
\end{document}